%% file: version6.tex
\documentclass[11pt]{article}
\input{YJ.tex}

\usepackage{amsmath}
\usepackage{amsfonts}
\usepackage{mathrsfs}
\usepackage{multirow} 
\usepackage{longtable}
\usepackage{geometry}
\usepackage{graphicx}
\usepackage{listings}
\usepackage{subfigure}
\usepackage{amssymb}
\usepackage{url}
\usepackage{xcolor} 
\usepackage[numbers, sort&compress]{natbib}
\usepackage{tabularx}
\usepackage{pdfpages}
\usepackage{emptypage}
\usepackage{subfigure}
\usepackage{float}
\usepackage{authblk}

\newcommand{\omg}{\omega_{0d}}

\title{Hi3+3: A Model-Assisted Dose-Finding Design Borrowing Historical   Data }

\author[1]{Yunshan Duan}
\author[2]{Sue-Jane Wang \footnote{This article reflects the views of the author and should not be construed to represent the views or policies of the U.S. Food and Drug Administration}}
\author[3, *]{Yuan Ji}
\affil[1]{{\small Department of Mathematics, Fudan University}}
\affil[2]{{\small Center for Drug Evaluation and Research, US FDA}}
\affil[3]{{\small Department of Public Health Sciences, University of Chicago}}
\affil[*]{{\small Corresponding author:  Yuan Ji, YJi@health.bsd.uchicago.edu}}

\begin{document}
\maketitle

\begin{abstract}

\noindent \textbf{Background --} 
In phase I clinical trials,  historical data may be available through multi-regional programs, reformulation of the same drug,   or previous trials for a drug under the same class.    Statistical designs that borrow information from historical data can reduce cost, speed up drug development, and maintain safety. \\
\textbf{Purpose --}
  Based on a   hybrid   design that partly uses probability models and partly uses algorithmic rules for decision making, we aim to improve the efficiency of the   dose-finding trials in the presence of historical data,   maintain safety for patients, and achieve a level of simplicity for practical applications.   \\ 
\textbf{Methods --}
We propose   the   Hi3+3 design, in which the letter ``H''   represents  ``historical data''.      We apply the idea in power prior to borrow historical data and   define the effective sample size (ESS) of the prior. Dose-finding decision rules follow the idea in the i3+3 design   \citep{liu2020i3+}     while incorporating    the historical data via the power prior and ESS.  The proposed Hi3+3 design  pretabulates the dosing decisions before the trial starts,   a desirable feature for ease of application in practice.    \\ 
\textbf{Results --}
The Hi3+3 design is superior than the i3+3 design due to information borrow from historical data. It is capable of maintaining a high level of   safety   for trial patients without sacrificing the ability to identify the correct MTD.  Illustration of this feature are found in the simulation results. \\
\textbf{Conclusion --}
With the demonstrated safety, efficiency, and simplicity,   the Hi3+3 design   could be a desirable choice    for dose-finding trials borrowing historical data.  

\end{abstract}

\section{Introduction}

Phase I trials are the first-in-human studies using human subjects to test the potential therapeutic effects of a new drug. The goal of a phase I trial is to establish the safety profile of the drug in a dose-dependent manner, and to identify   the maximum tolerated dose (MTD),     defined as the highest dose with a toxicity probability close to or not higher than a target toxicity probability, say $p_T = 0.3$. With the input of the $p_T$ value for precise definition of the MTD, standard statistical designs aim to identify the highest dose close to $p_T$. More recently, interval-based designs also require   the   prespecification of an equivalence interval (EI), e.g.,   $(p_T-\epsilon_1,p_T+\epsilon_2)$ for small fractions of $\epsilon_1$ and $\epsilon_2$. For example, $p_T=0.25$ and EI = $(0.2, 0.3)$.   The use of EI allows   doses with probability of toxicity within the EI to be considered as the MTD as well,   which can be elicited with trial clinicians.     Statistical   designs establish the MTD by assigning patients in an up-and-down manner over a grid of ascending dose levels. Typically, patients are enrolled in cohorts of certain size, say 3,   and sequential dose-allocation decisions continuously assign patient cohorts to appropriate doses according to the designs and the observed DLT outcomes.  



  In modern drug development,   historical data may be available for a drug in a future phase I dose-finding trials.   For example, through multi-regional drug developmental programs, a drug   that undergoes phase I trials in the Northern American region may be later tested in the Asian Pacific region with overlapping doses. Therefore, DLT data from   North America   may be borrowed to help improve the design and conduct of phase I trials in   Asia Pacific.   Other situations where historical data may be available include phase I trials using the same class of targeted therapies, or the same drug but with different formulation or tested in different disease indications.   A statistical design borrowing information from historical data may   increase the efficiency of the new trials, reduce their sample sizes, and enhance or maintain safety for trial patients. 


Various statistical designs for phase I trials  have been proposed and could be generally divided into two classes: rule-based designs and model-based designs.   For rule-based methods,   the   3+3 design \citep{storer1989design} and the i3+3 design   \citep{liu2020i3+}   both use prespecified decision rules to determine the dose for   trial patients.   Free of statistical models and   model-based   inferences, both designs are attractive to nonstatisticians and easy to implement in practice. However, i3+3 uses smarter rules that   are mindful of   variabilities in the observed data, is safer, and exhibits superior operating characteristics than the 3+3 design. 

Model-based designs,   such as the continuous reassessment method (CRM) \citep{o1990continual}, BLRM \citep{neuenschwander2008critical}, and mTPI \citep{ji2010modified}, 
use parametric curves or curve-free statistical models to describe dose-toxicity relationship.   With statistical modeling, these designs 
may naturally incorporate historical data through prior distributions. 
  For example, in \citep{Neuen2016codata} the BLRM design is combined with meta-analytic prior \citep{schmidli2014robust} to accommodate the ``Co-Data'' from historical or ongoing trials.  More recently,   the iCRM design \citep{zhou2020incorporating} proposes an   informative \hei ``skeleton'' for CRM to utilize historical information. Other related work based on CRM  that incorporates different ethnic and age groups can be seen in \cite{liu2015bridging} and \cite{li2020pa}. 

Interval-based designs \citep{ji2010modified,guo2017bayesian,yan2017keyboard,ivanova2007cumulative,liu2015bayesian} also use statistical models although the models do not assume a dose-response curve. Instead, toxicity outcomes and probabilities across different doses are assumed independent, and inference is centered around the EI and up-and-down decision rules. Recently,  a new framework has been proposed for   interval-based dose-finding designs   to incorporate historical data, which gives rise to the iBOIN,  iBOIN$_R$, and iKeyboard designs \citep{zhou2020incorporating}. The framework   considers   an    informative prior using skeletons similar to that in CRM   and builds on the concept of prior effective sample size (PESS) \citep{morita2008determining} for prior construction. 

  We propose the Hi3+3 design   
as a simple and safe method that incorporates historical trial data in a robust manner. Here, ``robust'' means that even when the historical data do not accurately reflect the true toxicity probabilities of the doses, the   proposed   design can still limit the chance of exposing patients from being exposed to overly toxic doses. This is an important practical feature in borrowing historical data for dose-finding designs, since the noise and variabilities in the historical data must be mitigated when they affect the safety of the designs.   To borrow information from historical data and still achieve robustness, the proposed Hi3+3 design resorts to a hybrid   
 framework that includes two components. Component I uses model-based inference to derive the dose-dependent effective sample size (DESS)   through a power prior \citep{ibrahim2015power}   and component II is a model-free dose-finding algorithm   following the same idea in   the i3+3 design. In other words, statistical models are used to assist the model-free decision making in determining DESS, therefore the term ``hybrid''.



The remainder of the paper is organized as follows. In Section 2, we introduce the power prior \citep{ibrahim2015power} and a model-free dose-finding algorithm.   Section 3 presents the pretabulated decision tables for Hi3+3,  a userful feature in practice.   Section 4 conducts simulations to assess the performance of Hi3+3 compared to i3+3. Section 5 briefly discusses the savings in sample size using Hi3+3, and section 6 ends the paper with a discussion.  Additional  theoretical justification and numerical results illustrating the performance of various designs are provided in the Appendix.  

An online R Shiny tool \url{https://hi3design.shinyapps.io/hi33_r_shiny/} 
is provided to illustrate the Hi3+3 design.

\section{Trial Design}

\subsection{Component I: Probability Model for Historical Data}
We use the power prior \citep{ibrahim2015power},   in the form of $p(y|\theta)^{\omega_0}\pi(\theta)$ to incorporate historical data from previous clinical trials. The power prior can be viewed as  a ``down-weighted  posterior'' based on an initial (noninformative) prior $\pi(\theta)$, a likelihood function of the historical data $p(y|\theta)$,   and a discounting factor $\omega_0$ that ``weakens'' the likelihood. The parameter $\omega_0$ represents the amount of information borrowed from the historical data, and usually ranges from 0 to 1. Specifically, $\omega_0 = 0$ means no information is borrowed whereas $\omega_0 = 1$ implies the historical data weights the same as the data to be obtained in the future trial, i.e., historical data is fully borrowed. A value between 0 and 1 means the historical data is down-weighted and only partial information is borrowed. 

Assume a subset of doses of the drug is overlapping in the previous and current trials, that is, the   doses $\{1,...,D\}$ in the current trial contain \hei some doses from the historical trial and others that are not. To be specific,   denote   the historical data $M_0= \{(x_{0d},n_{0d})| d=1,...,D\}$, where $d=1,...,D$ doses, in which $n_{0d}$ is the number of patients treated and $x_{0d}$ the number of patients experience DLTs at dose $d$. For the doses not used in previous trials, set $x_{0d}=n_{0d}=0$. \color{black}


For the historical data, 
a binomial likelihood function is constructed assuming $x_{0d}|p_d \sim bin(n_{0d},p_d)$. An initial vague prior $p_d \sim \pi_0(p_d)=beta(a_0,b_0)$ is assumed,   where $a_0 =b_0 =0.005$.   They represent a vague prior for the binomial likelihood with a tiny effective sample size of $0.01$   \citep{morita2008determining}.  

We propose a modified power prior   to borrow information   given by
\begin{align}
\pi(p_d|x_{0d},n_{0d},\omega_{0d}) &\propto  \prod_{d=1}^D p(x_{0d} | p_d)^{\omega_{0d}} \pi_0(p_d) \nonumber \\
                                   & = p_d^{\omega_{0d}x_{0d}} (1-p_d)^{\omega_{0d}(n_{0d}-x_{0d})} p_d^{a_0-1} (1-p_d)^{b_0-1} \nonumber \\
  & = p_d^{\omg x_{0d} + a_0 -1} (1-p_d)^{\omg (n_{0d}-x_{0d}) + b_0 -1}. \label{eq:power1}
\end{align}
Here $p(x_{0d} |p_d)$ is the binomial density of $x_{0d}$ and   $\omg \in [0,1]$   is the dose-dependent power parameter for dose $d$.  \citet{duan2006evaluating}   noted that the construction in \eqref{eq:power1} violates the likelihood principle if the power parameters $\omg$ are treated as random quantity and assigned a prior distribution. Therefore, a proper construction   with   a prior on $\omg$   requires inclusion of the normalizing constant $ \int p(x_{0d}|p_d)   ^{\omega_{0d}}   \pi_0(p_d) d p_d$ as a function of the power parameter. However, we resort to a different and simpler approach. We fix $w_{0d}$ at a value that is expected to induce desirable safe decisions for patients in the trial but also allow maximum borrowing. This will be explained in Section 2.3. Because $w_{0d}$'s are fixed in our construction, the modified power prior \eqref{eq:power1}     does   not violate the likelihood principle.   

It is immediate that \eqref{eq:power1} follows a beta distribution, i.e.,
\begin{align}
p_d|x_{0d},n_{0d},\omega_{0d} & \sim beta(\omega_{0d}x_{0d}+a_0,\omega_{0d}(n_{0d}-x_{0d})+b_0)  \nonumber \\
& \equiv beta(a_d,b_d), \label{eq:pp0}
\end{align}
with $a_d= w_{0d}x_{0d} +a_0$ and $b_d= \omega_{0d}(n_{0d}-x_{0d})+b_0$, for $d=1, \ldots, D.$ The power prior mean of $p_d$ is $\hat{p}_d = a_d/(a_d + b_d).$ The power parameter $\omega_{0d}$   weighs   the historical   information   relative to the   information in the current trial data,   
and thus controls the influence of the historical data on the inference of the current trial, in a dose-dependent fashion.



  Using $\omg,$ we quantify the 
amount of information contained in the prior   by the effective sample size (ESS). Again,   we resort to  \cite{morita2008determining} which argues that    a beta prior distribution $beta(a,b)$ has   ESS    $(a+b)$ for a binomial likelihood. Therefore, for each dose $d$, the dose-dependent effective sample size (DESS)   is
\begin{align*}
m_d &= a_d + b_d= a_0+ b_0 + \omega_{0d} n_{0d}, \; d=1,...,D.
\end{align*}
The DESS quantifies the amount of borrowed information and will be used for dose-finding rules later. However, in order to enforce the monotonic assumption of dose-toxicity relationship, we take one more action on the power prior: we perform isotonic regression \citep{robertson1988order} on the power prior mean $\hat{p}_d$ and denote the transformed mean $p^\star_d$. As a result, the corresponding isotonic-transformed power prior can be denoted as
\begin{equation}
p_d|x_{0d}, n_{0d}, w_{0d} \sim beta(a_d^\star, m_d - a_d^\star) \label{eq:power2}
\end{equation}
where $a_d^\star = m_d \cdot p_d^\star$ is the isotonic-transformed parameter induced by $p_d^\star$. It could be interpreted as the prior expected number of DLT   since $E(p_d)=a_d^\star / m_d$ based on \eqref{eq:power2}.   This interpretation will be exploited in Component II next. Importantly,   the DESS of \eqref{eq:power2} is still $m_d$, i.e. the isotonic transformation does not change the   DESS.   
The isotonic-transformed power prior means $p^\star_d$'s along with the DESS $m_d$'s will be used for Component II of the Hi3+3 design, explained next. 

\paragraph*{An example} To illustrate the steps in Component I, consider historical trial data for a completed dose-finding trial with $D=5$ doses   and   observed outcome $\{(x_{0d}, n_{0d})\}_{d=1}^5 = \{(1,6), (0, 3), (0, 3), (2,6), (3,3)\}.$ Suppose $w_{0d} = 1$ for illustrative purpose. According to \eqref{eq:power1}, before the isotonic transformation, the power prior   \eqref{eq:pp0}   for each of the five doses is $beta(1.5, 5.5), \; beta(0.5, 3.5), \; beta(0.5, 3.5), \; beta(2.5, 4.5), \; beta(3.5, 0.5),$ respectively. The   corresponding   power prior means are $\hat{p}_1 = 0.214, \; \hat{p}_2 = 0.125, \;\hat{p}_3 = 0.125, \;\hat{p}_4 = 0.357, \;\hat{p}_5 = 0.875$ and the DESS are $m_1 = 7, \; m_2=4, \; m_3=4, \; m_4=7, \; m_5=4$ for the five doses. After the isotonic transformation   using the pool adjacent violator algorithm \citep{bacchetti1989additive},   the isotonic-transformed power prior means become $p_1^\star=0.154, \; p_2^\star=0.154, \; p_3^\star=0.154, \; p_4^\star=0.357, \; p_5^\star=0.875.$ Keeping the same DESS $\{m_d\}_{d=1}^5$, the isotonic-transformed power prior \eqref{eq:power2} for the five doses become $$beta(1.078, 5.922), \; beta(0.616, 3.384), \; beta(0.616, 3.384), \; beta(2.5, 4.5), \; beta(3.5, 0.5), $$ with prior expected number of DLTs $m_d \times p_d^\star$ given by, $$a_1^\star=1.078, \; a_2^\star=0.616, \; a_3^\star=0.616, \; a_4^\star=2.5, \; a_5^\star=3.5.$$

\subsection{Component II: Dose-finding algorithm} \label{sec:df}

  We apply the principles in the i3+3 design and utilize the power prior \eqref{eq:power2} for decision making.   
Let $p_T$ (e.g. $p_T=0.3$) denote the target probability of toxicity   for the true MTD    and the equivalent interval be EI$=(p_T-\epsilon_1,p_T+\epsilon_2)$ (e.g. EI=$(0.25,0.35)$, with $\epsilon_1=\epsilon_2=0.05$). Suppose dose $d$ is the current dose administered to enrolled patients, $n_{d}$ the number of patients that have been treated, and $x_{d}$ the number of patients that have experienced DLTs at dose $d$, respectively.   Denote the current trial data by $M=\{(x_d,n_d), d=1,...,D\}$. Also, denote $ W=\{\omega_{0d}, d=1,...,D\}$ the set of dose dependent power parameters. Finally, recall that $M_0=\{(x_{0d},n_{0d}), d=1,...,D\}$ denotes the historical data on the $D$ doses,   with $n_{0d}=0$ for doses not in the historical data.

Below, we first present the decision rules for dose finding using Hi3+3, and explain the logic behind. At the current dose $d$, calculate two quantities $\displaystyle ( \frac{x_d+a^\star_d}{n_d+m_d},\frac{x_d+a^\star_d-1}{n_d+m_d}).$ \bigskip

{\small
\noindent \fbox{%
  \parbox{0.99\textwidth}{
    Algorithm 1: 
          \begin{itemize}
            \item If $\displaystyle \frac{x_d+a^\star_d}{n_d+m_d}$ is below the EI, escalate ``E'' and enroll patients at the next higher dose $(d+1)$; 
		
            \item Else, if $\displaystyle \frac{x_d+a^\star_d}{n_d+m_d}$ is in the EI, stay ``S'' and continue to enroll patients at current dose $d$;
		
            \item Else, if $\displaystyle \frac{x_d+a^\star_d}{n_d+m_d}$ is above the EI, there are two scenarios:
              \begin{enumerate}
              \item	If $\displaystyle \frac{x_d+a^\star_d-1}{n_d+m_d}$ is below the EI, stay ``S'' and continue to enroll patients at the current dose $d$;
		
              \item	Else, de-escalate ``D'' and enroll patients at the next lower dose $(d-1)$.
              \end{enumerate}
            \end{itemize}
                }
}
}

The decisions rules above follow the idea in the i3+3 design, which use $(\frac{x_d}{n_d}, \frac{x_d-1}{n_d})$ instead of the two fractions $ ( \frac{x_d+a^\star_d}{n_d+m_d},\frac{x_d+a^\star_d-1}{n_d+m_d}).$  in algorithm 1.  The two new fractions are used to incorporate historical data. Recall that the power prior in \eqref{eq:power2} for a dose $d$ implies that $m_d$ patients have been treated and an expected number of $a_d^\star$ patients have DLTs in the historical data.   
Therefore, Hi3+3 combines  the historical data and the current trial data in the following way:   1) consider $m_d$ as the effective ``number'' of treated patients and $a_d^\star$ the effective ``number'' of patients having DLT in the historical data, and 2) combining them with current data $n_d$ and $x_d$, we arrive at 
a combined   total of $(n_d+m_d)$ treated patients and $(x_d+a_d^\star)$ patients experienced DLTs at dose $d$ . Based on this intuition, the Hi3+3 dose-finding algorithm   follows directly i3+3, as shown in Algorithm 1.

\paragraph*{Safety rules} Similar to the i3+3 design, we add two safety rules as ethical constraints to avoid excessive toxicity: 

\begin{itemize}
  \item[] {\bf Safety rule 1 (early termination):}  Suppose that the lowest dose, dose 1, has been used to treat patients, i.e., $n_1 >0$. If $Pr(p_1>p_T|M_0, M, W)>\xi$ for $\xi=0.95$, terminate the trial due to excessive toxicity.  

  \item[] {\bf Safety rule 2 (dose exclusion):} For any dose $d$, if $n_d>0$ and $Pr(p_{d}>p_T|M_0, M, W)>\xi$ for $\xi=0.95$, then   dose $d$ and higher doses are removed from the trial. \hei Any future escalation to dose $d$ will be changed to ``S'', stay.
\end{itemize}

  The posterior distribution in both safety rules is based on the combined data from the current trial and the historical trial.   In particular, we use $beta(x_d + a_d^\star + 1 - a_0, n_d - x_d + m_d - a_d^\star + 1 - b_0)$ that corresponds to a posterior based on an isotonic-transformed power prior for the historical data and the binomial likelihood for the current trial data.   Appendix A   gives an explanation for the use of this beta posterior distribution.   

  Lastly, under the Hi3+3 design, a trial is terminated either when a prespecified sample size is reached or according to { Safety rule 1.}

\subsection{Determination of $\omega_{0d}$} \label{sec:det-ess}

  In the isotonic transformed power prior, the DESS   for dose $d$ is controlled by $\omega_{0d}$, which represents the amount of information borrowed from historical data   for each dose.   Ideally, one should borrow more if the historical data at dose $d$ is ``compatible'' with the true toxicity probability of the dose. For example, if the true probability is 0.3, and the historical data is $(x_{0d}, n_{0d}) = (1, 3)$, the data is compatible.   It is not compatible if $(x_{0d},n_{od})= (3,3), (2, 3),$ or $(0,3)$.   However, in practice the true toxicity probabilities are unknown and therefore it is not   possible to identify   which dose has compatible historical data.   Therefore, researchers tend to focus on the determination of ESS through practical considerations.   
 For example, 
the iBOIN design \citep{zhou2020incorporating} uses a fixed ESS $\in [1/3(N/D),1/2(N/D)]$ where $N$ is the maximum sample size and   $D$ \hei is the number of doses for the current trial. 
  We take a different approach   
to determine $\omg$ and DESS based on the compatibility between the historical data and the expected data from the current trial. The proposed approach   aims to ensure the safety of the design, limiting the frequency of aggressive decisions when borrowing information from historical data. 


The Hi3+3 dose-finding Algorithm 1 in Section \ref{sec:df} can be thought of as the i3+3 design using the pseudo data $(x_d + a_d^\star, n_d + m_d)$ where $(a_d^\star, m_d)$ are the pseudo number of patients with DLTs and pseudo number of patients treated at dose $d$ from the historical data. Both quantities $(a_d^\star, m_d)$ are functions of $\omg$ which control the amount of information borrowed. If no information is borrowed, the algorithm reduces to the i3+3 design based on the current trial data $(x_d, n_d)$ at dose $d$. Specifically, we propose the following three conditions to help determine the power parameters $\omega_{0d}$'s.   We first define that   decision ``A''    is a more aggressive decision than ``B'' if ``A'' leads to a higher dose level for future patients than ``B''. For example, ``E'' -- escalate to the next higher dose is more aggressive than ``S'' -- stay ("S") at the current dose, which in turn is more aggressive than ``D'' -- de-escalate to the next lower dose.  

\bigskip

\noindent
{\bf Condition 1 ($\alpha$-Tolerability):  } {\it A value of $\omg$ is said to satisfy the   $\alpha$-Tolerability   condition if for a maximum number of patients $n_d$ at a dose $d$, say $n_d = 15$, the Hi3+3 decisions that borrow from historical data (i.e., based on $(x_d + a_d^\star, n_d + m_d)$ ) contain no more than $\alpha$ proportion of {more aggressive decisions} than the i3+3 decisions that do not borrow (i.e., based on $(x_d, n_d)$). Here, $\alpha$ is a tuning parameter that is specified by the trial designer, e.g. $\alpha= 0.1$.}

\bigskip


\bigskip

\noindent
{\bf Condition 2 (K-Ceiling):} {\it For fixed $a_0$ and $b_0$, a value of $\omg$ is said to satisfy K-Ceiling condition if it does not lead to $m_d > K$, where $K$ is a prespecified threshold (say, $K=9$). Note $m_d =a_0 + b_0 +\omega_{0d} \cdot n_{0d}$. }

\bigskip

\noindent Condition 2 
restricts the historical information from having too much effect on the current trial,   by limiting the value of DESS to be no   more than $K$, where $K$ (e.g. $K=9$) is a threshold specified in advance.

\bigskip

\noindent
{\bf Condition 3 (Retaining):} {\it A value of $\omg$ must not lead to exclusion of a dose in the trial based on the historical data $M_0$ alone. This means $Pr(p_d>p_T| M_0, W)$, calculated under $beta(a_d^\star +1-a_0, m_d-a_d^\star +1-b_0)$, must be less than $\xi$, say $\xi=0.95$. }

\bigskip

\noindent This condition states that a candidate dose included in the current trial must not be deemed unacceptable and overly toxic under the isotonic transformed power prior.

\bigskip

To calculate the maximum $(\omega_{01},...,\omega_{0D})$ that satisfies the three conditions, we need to solve a multidimensional constrained optimization problem that involves a nonlinear isotonic transformation of functions of $\omega_{0d}$'s   (for Condition 1).   This is complex and time-consuming. Instead, we propose an effective computational algorithm to determine $\omg$'s   that satisfy all three conditions.   
The algorithm essentially approximates the true solution of the optimization problem by $1)$ isotonically transforming the historical data, $2)$ using bisection to find the $\omega_{0d}$'s that satisfy the three conditions under the transformed historical data, which is much easier, and $3)$ isotonically transforming the implied power prior in $2)$ again and solves   for   the final $\omega_{0d}$'s.

\bigskip

{\small
\noindent \fbox{%
  \parbox{0.99\textwidth}{ Algorithm 2: 

\paragraph*{Step 1: Generate pseudo historical data by conducting isotonic transformation on the historical data.} Denote the historical data $M_0= \{(x_{0d},n_{0d})| d=1,...,D\}$ and denote $\tilde{D}=\{d|n_{0d} \neq 0, d=1,...,D\}$. Let $p_{0d}= x_{0d}/n_{0d}$ when $d \in \tilde{D}$.  For $(d_1,...,d_k) \in \tilde{D}$,   denote   $(p'_{0d_1},...,p'_{0d_k})$ as the isotonic transformation of $(p_{0d_1},...,p_{0d_k})$. For $ d \notin \tilde{D}$, $p'_{0d}$ is randomly generated from the uniform distribution between the $p'_0$s of the adjacent lower and higher doses of $d$ in $\tilde{D}$. Let $x'_{0d}= n_{0d}p'_{0d}$ and $M'_0= \{(x'_{0d},n_{0d})| d=1,...,D\}$ is the pseudo historical data.
	
\paragraph*{Step 2: Determine an intermediate set of $(\omega'_{01},...,\omega'_{0D})$   using bisection to ensure the three conditions are satisfied based on the pseudo historical data. }		Denote the initial prior as $beta(a_0,b_0)$. Given the pseudo historical data from step 1, the power prior is $beta(a'_d,b'_d)  \equiv beta(x'_{0d}\omega'_{0d}+a_0, (n_{0d}-x'_{0d})\omega'_{0d}+b_0)$. Under this pseudo prior in Hi3+3, use bisection to calculate the maximum $\omega'_{0d}$ on each dose $d$ that satisfies the three conditions.   Under the maximum $\omega'_{0d}$,   the estimated prior mean is $p'_d=a'_d/(a'_d+b'_d)$ which may not be monotonic across doses.
	
\paragraph*{Step 3: Conduct isotonic transformation on the prior means $(p'_1,...,p'_D)$ to obtain a monotonic vector $(p^*_1,...,p^*_D)$ and reversely solve the corresponding weight $(\omega^*_{01},...,\omega^*_{0D})$.}
	Let $(p^*_1,...,p^*_D)$ denote the isotonic transformed $(p'_1,...,p'_D)$. We want to obtain a prior $beta(a^*_d,b^*_d) \equiv beta(x'_{0d}\omega^*_{0d}+a_0, (n_{0d}-x'_{0d})\omega^*_{0d}+b_0)$ where $a^*_d/(a^*_d+b^*_d)=p^*_d$ for $d=1,...,D$. Therefore,
	\begin{equation*}
	\omega^*_{0d}=\begin{cases}
	
	\frac{a_0-p^*_d (a_0+b_0)}{p^*_d n_{0d}-x'_{0d}} & d \in \tilde{D}\\
	
	1 & d \notin \tilde{D}
	
	\end{cases}
	\end{equation*}
}}}

\bigskip

  Theoretically, there is no guarantee that the $\omega^*_{0d}$'s would satisfy the three conditions; only the $\omega'_{0d}$'s in step 2   would.   But for pseudo historical data, however, we find that in practice for small phase I trial data usually with no more than a dozen of patients per dose, the algorithm works remarkably well. For example, using massive simulations with randomly generated historical data, the $\omega^*_{0d}$'s obtained by Alogorithm 2 all satisfy the three conditions. Specifically, in a dataset of 1,000 random historical trials, Algorithm 2 generate DESS's that lead to fewer than $\alpha$ proportion of more aggressive decisions  ($\alpha$-Tolerability),    are no larger than $K$ (K-Ceiling),   and retain all doses based on the power prior (Retaining).


\subsection{Estimation of MTD}
Hi3+3 uses the same procedure as i3+3 to estimate the MTD at the end of the trial, which is based on isotonic-transformed posterior mean of $p_d$   under the   binomial likelihood for the observed data $\{(x_d,n_d)\}$ and a prior for $p_d$.   We first decide a set of candidate MTD doses based on inference using two priors for $p_d$:   {\it prior 1)} the power prior $beta(a_d^\star,m_d-a_d^\star)$ in \eqref{eq:power2} and {\it prior 2)} an   initial vague prior $beta(a_0, b_0)$ where $a_0=b_0=0.005$. \hei Under each prior, we obtain the posterior means of  the corresponding posterior beta distributions, and apply the isotonic transformation   on the posterior means   to obtain   the transformed posterior means   $\tilde{p}_{1,d}$ for {\it prior 1)} and $\tilde{p}_{2,d}$ for {\it prior 2)}.  Then define a set of candidate doses for MTD as 
\begin{align*}
\mathcal{D}_{safe} = \{d|n_d>0 , \;\; \tilde{p}_{1,d} \leq p_T +\epsilon_2 \; or \; \tilde{p}_{2,d} \leq p_T +\epsilon_2 \}.
\end{align*}
In words, $\mathcal{D}_{safe}$ contains a set of doses that has been used in the trial with an isotonic-transformed posterior mean less than the upper bound of the EI based on either of the two priors. We found through simulation studies that   selecting MTD from $\mathcal{D}_{safe}$ using two priors   f
improves the   selection accuracy of Hi3+3   compared to from a set using only one prior, say the power prior. 

Lastly, the estimated MTD $d^*$ is the dose in $\mathcal{D}_{safe}$ with the smallest difference between $\tilde{p}_{1,d}$ and the target $p_T$, i.e., 
\begin{align*}
d^* = \mathop{\arg\min}_{d\in \mathcal{D}_{safe}} |\tilde{p}_{1,d}-p_T|.
\end{align*}

In words, the final MTD $d^*$ has the smallest distance between $\tilde{p}_{1,d}$ and the target $p_T$. 
If more than one dose of $d^*$ exists, it means there are doses for which the value $\tilde{p}_{1,d}$'s are the same. Then we choose MTD according to the following simple rules following i3+3:  
\begin{itemize}
\item[] a) If $\tilde{p}_{1,d^*} > p_T$, the estimated MTD is the lowest dose in $d^*$. 
\item[] b) If $\tilde{p}_{1,d^*} \leq p_T$, the estimated MTD is the highest dose in $d^*$. 
\end{itemize}

\section{Pretabulated Decision Tables}
An important feature of the Hi3+3 design is the pretabulated decision tables before the trial starts. This feature allows the investigators to examine, calibrate, and agree on the dosing decisions during the designing stage of the trial. In the interval designs such as the i3+3 design, a single decision table provides the decisions across all the doses. However, in the Hi3+3 design, due to the use of DESS, a separate decision table is needed for each dose. In particular, the available decisions are \{D, S, E\}, representing the decisions to de-escalate to the next lower dose, stay at the same dose, and escalate to the next higher dose, respectively.   These decisions can be denoted as $\mathcal{A}_d(M_0, M, W) \in $ \{D, S, E\}, a mapping from the historical data $M_0$, the observed data $M$, and DESS $W$ to one of the three decisions. 

\begin{figure}[htbp]
  \begin{center}
  \begin{tabular}{ll}
			\includegraphics[width=0.33\textwidth]{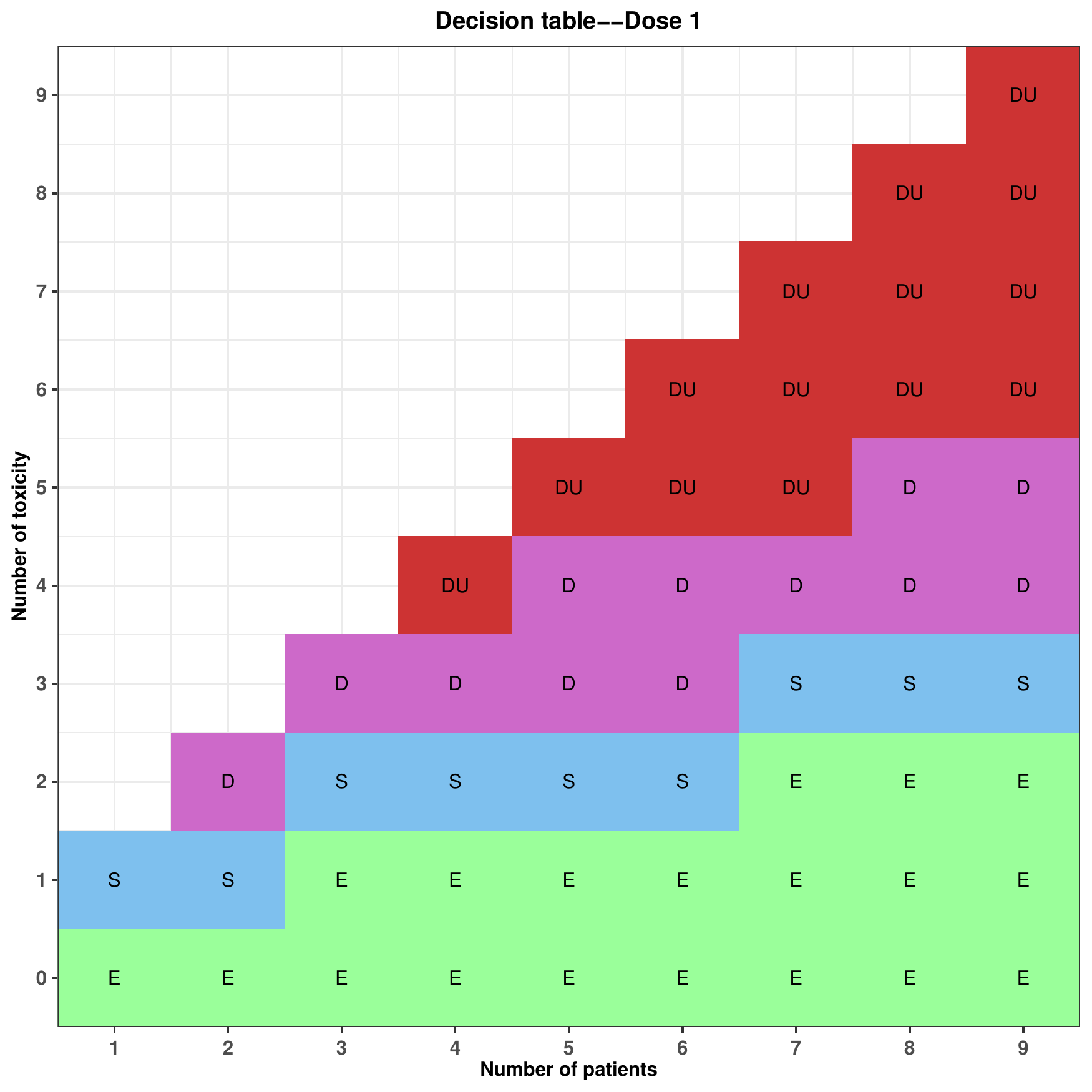}
			\includegraphics[width=0.33\textwidth]{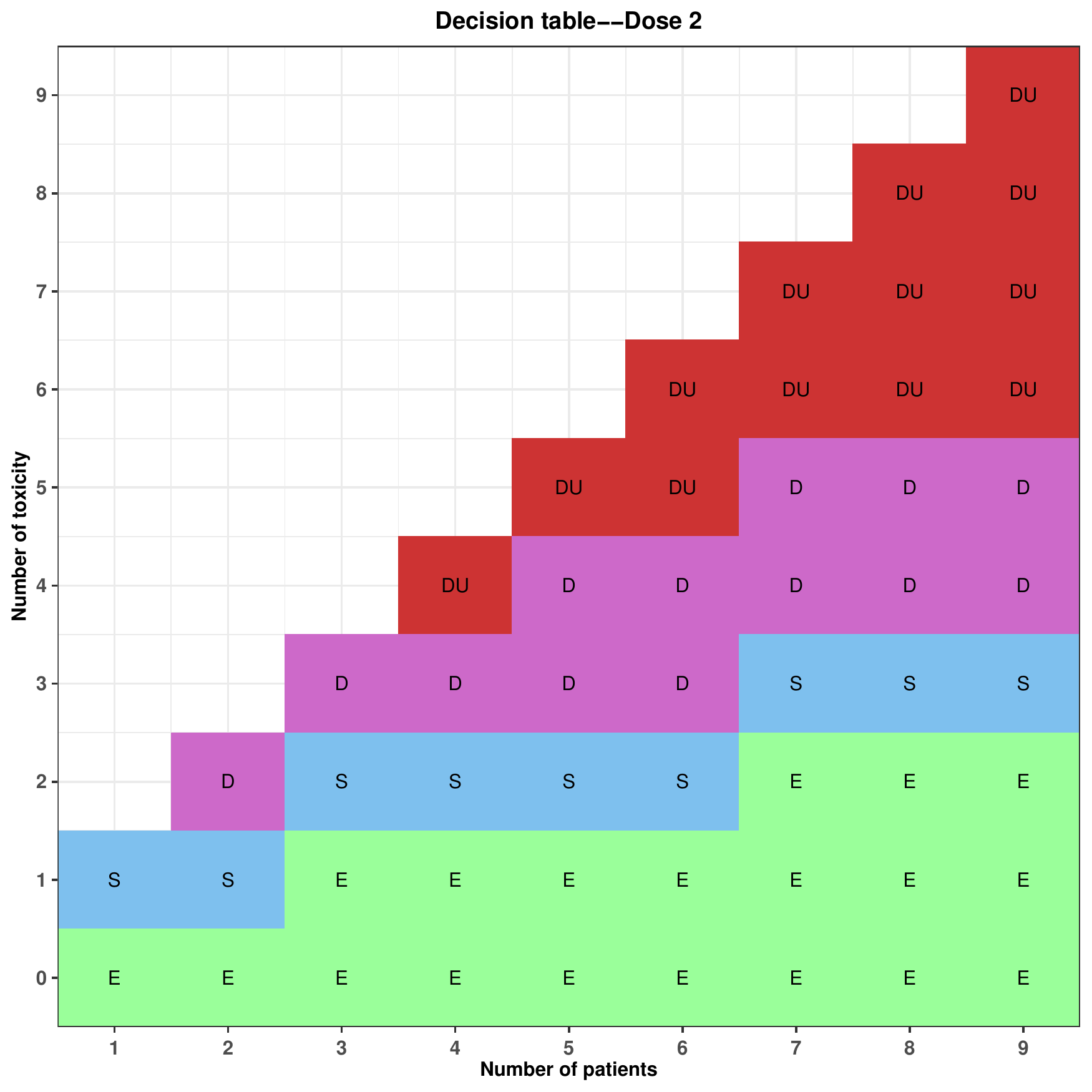}

			\includegraphics[width=0.33\textwidth]{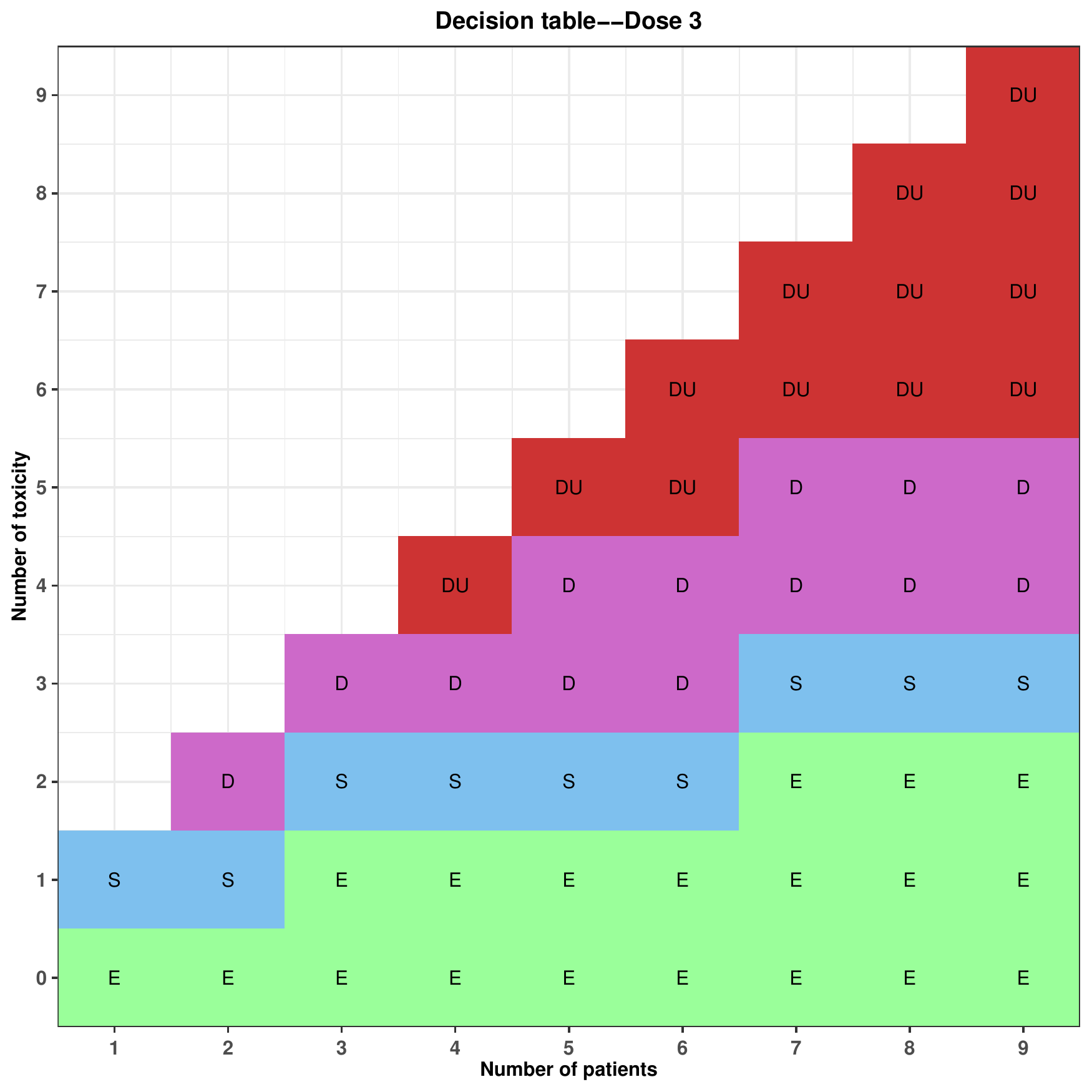} \\
			\includegraphics[width=0.33\textwidth]{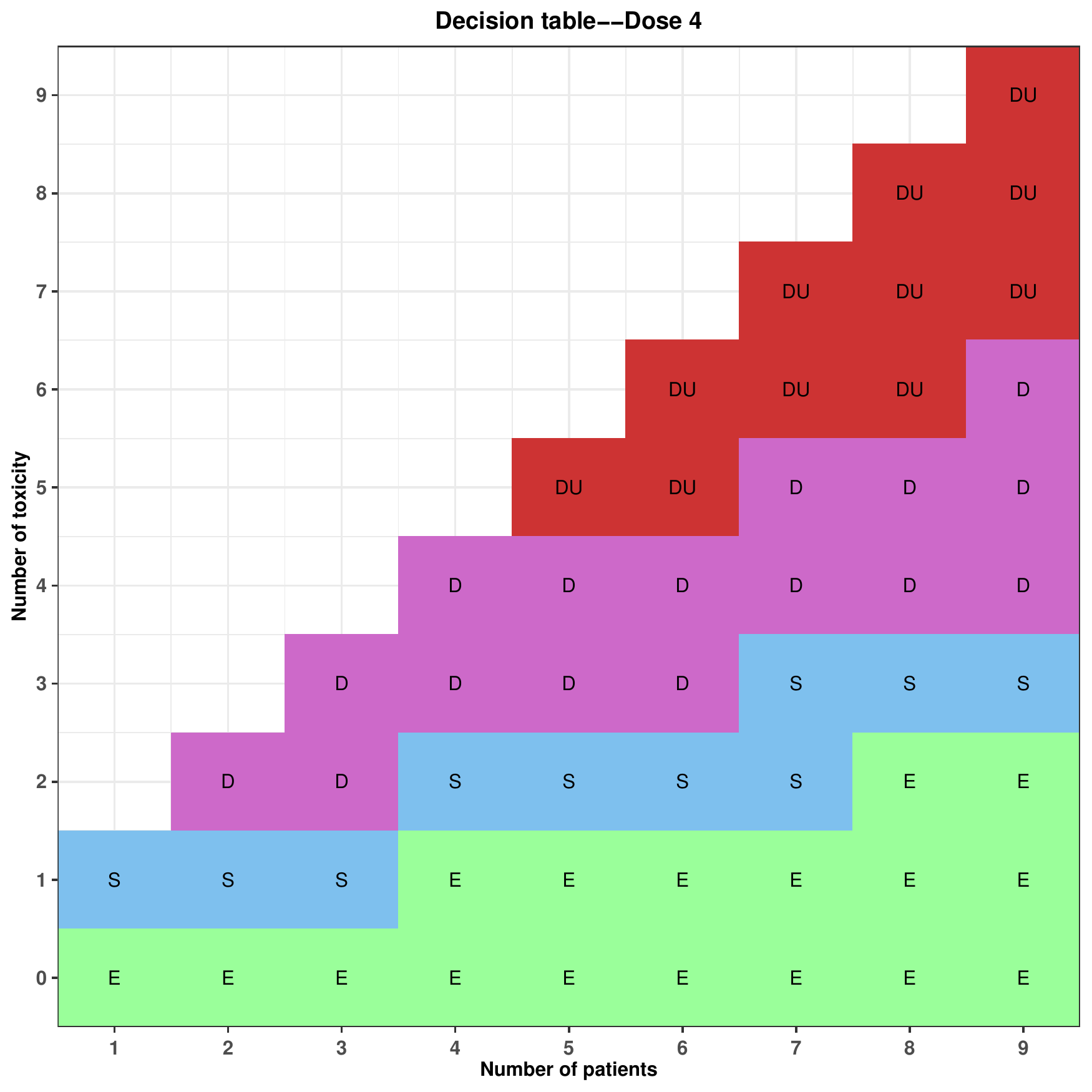}

			\includegraphics[width=0.33\textwidth]{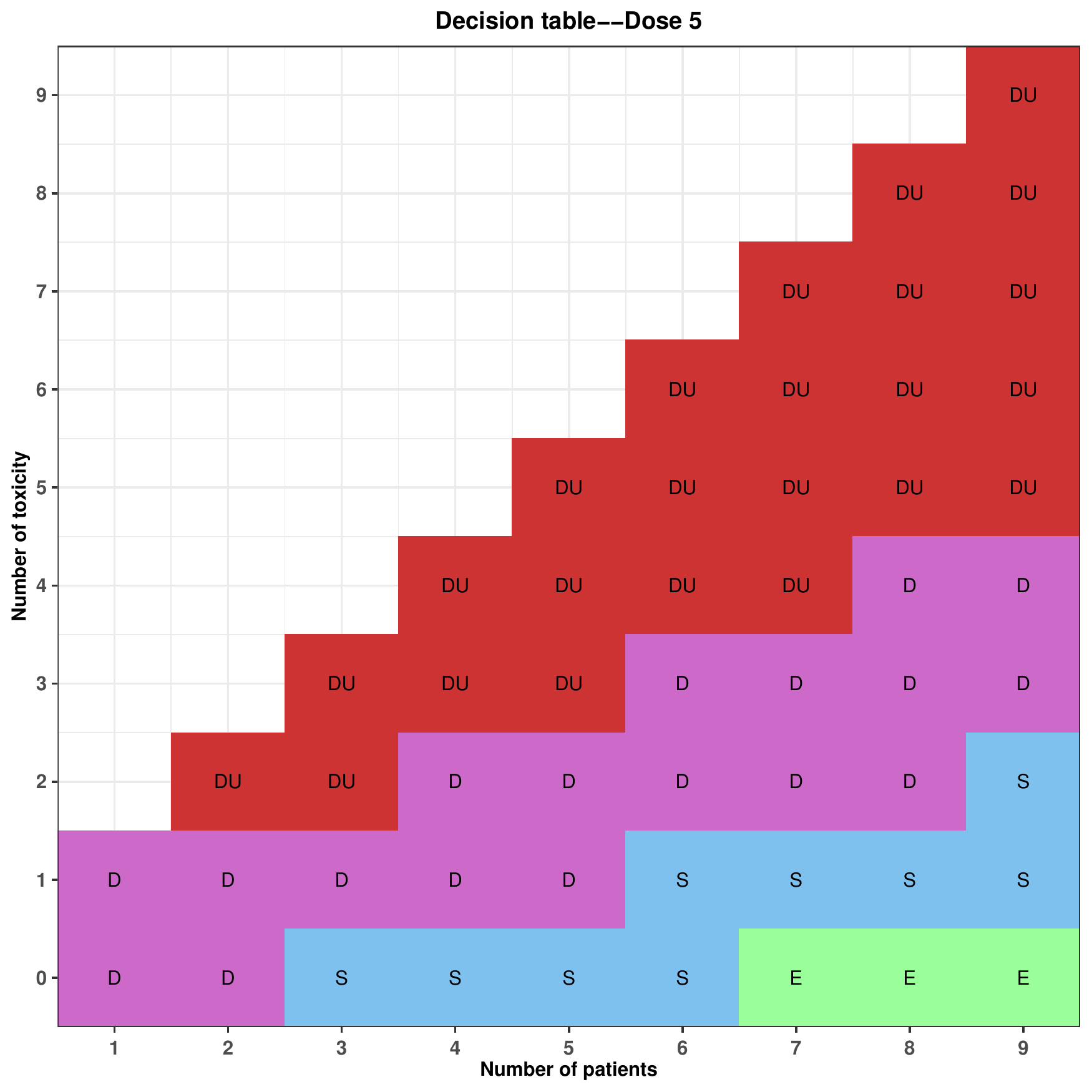}

  \end{tabular}
  \end{center}
	\caption{  Left right and top down are five decision tables of five doses under the Hi3+3 design,   with historical data $\{(0,3),(1,6),(1,6),(2,9),(3,6)\}$ and $\omega_0=(0.59,0.50,0.51,0.57,1)$ for the five doses.   The decision tables become more ``conservative'' with more ``D'''s and ``DU'''s and less ``E'''s, as the historical data on each dose show more toxicity.   }
	\label{figure: decision tables}
\end{figure}

An example is presented in Figure \ref{figure: decision tables}, which includes a fourth letter ``U'' indicating that the dose is unacceptable due to toxicity and is removed based on our Safety rule 2. Five decision tables of the five doses in Scenario 4 used in the simulation study in Section 4.2 are displayed. 

\hei

\section{Numerical   Examples }

\subsection{Simulation Study}

Simulation studies with comparisons and sensitivity analyses   are conducted to assess the performance of Hi3+3. For comparison, we include the i3+3 design as a comparison to illustrate the impact of borrowing information from historical data. Comparison to a couple of other designs, iBOIN and iBOIN$_R$,  also borrowing information is provided in Appendix B. 

\paragraph{Fixed scenarios} 

We assume $D=5$ doses and the target DLT probability $p_T = 0.3$. Both simulated trial data and historical data are generated based on different scenarios that specify the true dose toxicity probabilities. For the simulated trials, a sample size of 30 patients is used and patients enroll with a cohort size of 3. For the Hi3+3 and i3+3 designs, the EI is $(p_T-0.05, p_T+0.05)$. For Hi3+3, we set   $\alpha=0.1$ \hei and $K=9$. For the iBOIN design, we follow \citet{zhou2020incorporating} and use the isotonic pseudo prior as the skeleton and set the ESS of each dose as 3.   For the iBOIN$_R$ design, given the skeleton with dose $d^*$ as the prior estimator of the MTD, if $d^* > D/2$, the ESS is set as $(n_{01},...,n_{0d^*},0,...,0)$.   Under each simulation scenario, we generate a set of historical data $\{x_{0d}, n_{0d}\}$ for all five doses, and generate 10,000 simulated trials   with the same historical data.   The 
historical trial data is generated using arbitrary true toxicity probabilities, with a historical trial sample size of 30.

We use a total of 13 scenarios (Figure \ref{table:fixed_scenarios}) that represent a variety of dose-toxicity relationships. In the first   nine \hei scenarios, the current trials use exactly the same doses as the historical trials,   and in the last four scenarios, some new doses are included in the current trials.   Also,   in the first five scenarios, the historical data   is compatible with the true toxicity probabilities   and therefore, when borrowed, would improve the design and inference for the current trial. Here,   ``being compatible'' means   that the observed toxicity rates are similar to the true toxicity probabilities at various doses. Specifically, if a reasonable design is used to analyze the data, the dose corresponding to the true MTD would be identified as the estimated MTD   using   the historical data. 
In scenarios 6-9,  the observed data severely deviate from the true toxicity probabilities. Therefore, borrowing from the historical data from these scenarios  may not improve the design or inference for the current trial, since they do not reflect the true toxicity probabilities of the doses. In other words, the historical information is inaccurate   in these scenarios. \hei
In scenarios 10-13,   three doses are used in historical data and two new doses are included in the current trials.   In scenarios 10 and 11, new doses are extrapolated and added above the highest dose from the historical data and in scenarios 13 and 14, new doses are interpolated and inserted between the existing doses.

\begin{figure}[H]
	\centering
	\includegraphics[scale=0.5]{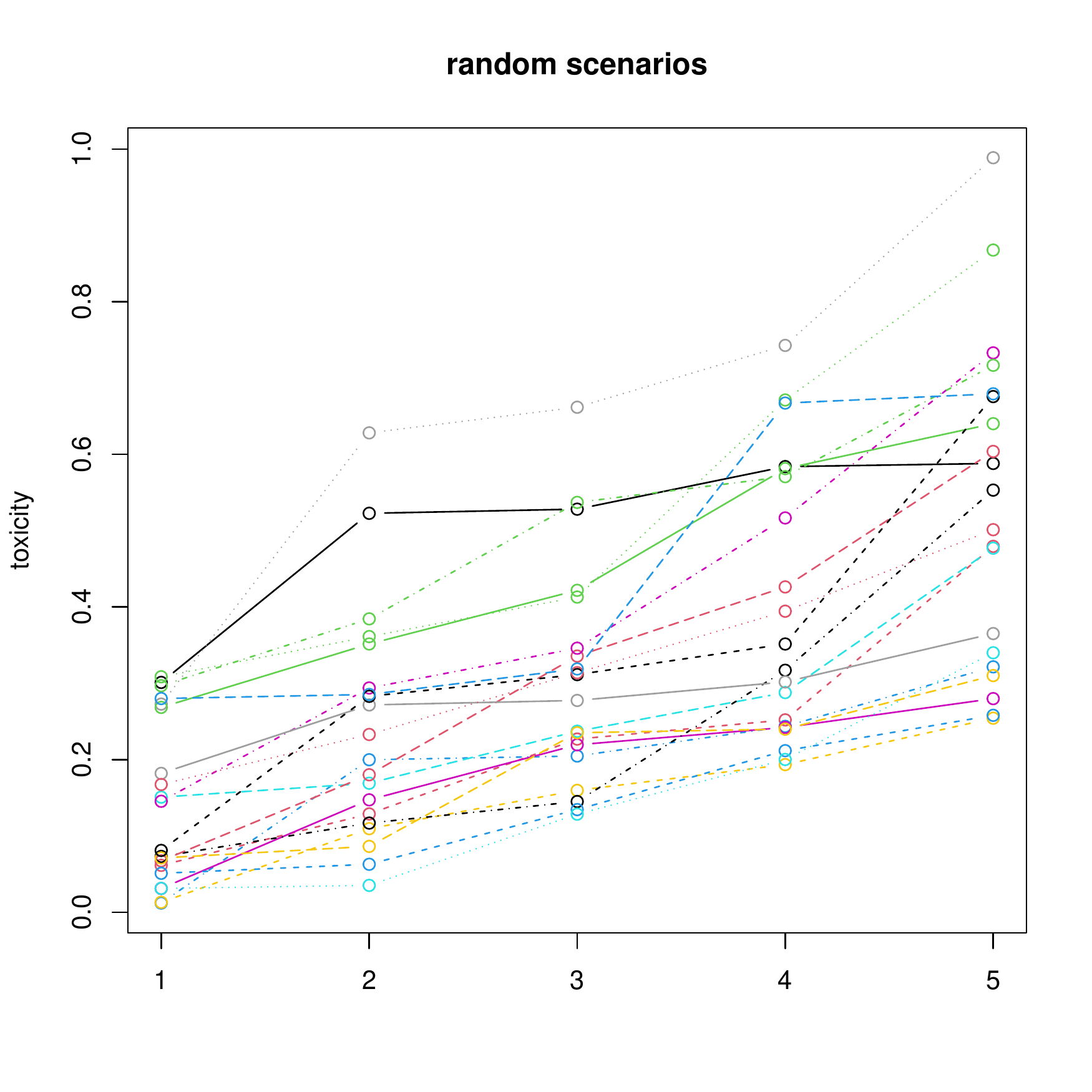}
	\caption{  Illustration of the first 20 random scenarios of true toxicity probabilities in the simulation.  }
	\label{fig:ran_scenarios}
\end{figure}

\paragraph*{Random scenarios} To corroborate the conclusions obtained from the above simulations in fixed scenarios, we further generated a large number of   1,000 \hei random scenarios to evaluate the overall performances. Scenarios are generated based on the pseudo-uniform algorithm \citep{clertant2017semiparametric}. The first set of 20 scenarios are plotted in Figure \ref{fig:ran_scenarios}. 
Generation of the historical data is the same as in the fixed scenarios.

\subsection{Simulation Results}
\paragraph{Fixed scenarios} We evaluate the performance of the designs based on several metrics   assessing the reliability (in terms of the ability to identify the MTD) and safety.   The simulation results for the 13 fixed scenarios are presented in   Table \ref{table:fixed_scenarios}   and Table \ref{table:fixed_scenarios_appendix} in Appendix B.   We provide a summary of results in Table \ref{table:fixed_scenarios} next, focusing on the comparison to i3+3 and the benefits of borrowing information using power prior and DESS.   A desirable design should demonstrate a good balance between patient safety and ability to identify the true MTD. 

When the historical data is compatible with the true toxicity probabilities, favorable prior distributions can be elicited by   the Hi3+3 design,   which in turn improve   its performance   in all the metrics. This is seen for scenarios 1-5 in which the i3+3 design is no match for Hi3+3. For example, Hi3+3 yields higher PCS (probability of correct selection of MTD), lower percentages of patients treated over MTD, and lower percentages of selecting dose over MTD than i3+3. 

In scenarios 6-9, the historical data leads to mis-specified priors. Interestingly, Hi3+3   is robust and does not exhibit   reduced performance when compared with i3+3. 

Scenarios 10 and 11 examine the cases that new and higher doses are   extrapolated and added   to the trial, at which no historical data are available. Hi3+3 assumes the historical data for the newly added doses has no information ($n_{0d}=0$ for the added doses). The performance of Hi3+3 is driven by the observed historical data on existing doses.   For example, in scenario 10, historical data at dose 3 suggests that the dose is safe and Hi3+3 tends to escalate more often than i3+3; in contrast, scenario 11 presents an opposite case where Hi3+3 is less likely to escalate to new doses.   

Scenarios 12 and 13 include interpolated doses and the Hi3+3 design shows similar traits in its performance, which are largely driven by the historical data. In general, the design does well when compared to the i3+3 design.

\paragraph*{Random scenarios}   Table \ref{table:random_result} shows a few operating characteristics for the random scenarios.   In a large scale,   Hi3+3 shows higher probabilities of selecting the true MTD and assigns fewer patients to doses over the MTD (i.e., safer) than i3+3.   

\input{code/table/result_random_table2}

\subsection{Sensitivity Analysis}
Sensitivity analysis is conducted to   further   evaluate the Hi3+3 design. First, we investigate the effect of the parameter $\alpha$, which impacts the decision tables of the Hi3+3 design. The larger its value, the larger $\omega_{0d}$ and more information borrowed. Using scenarios 1-5, we conduct simulations by setting $\alpha= 0.05,0.10,0.10,0.12,0.15$, and $0.20$ respectively. Table \ref{table:sen_tol} shows that as $\alpha$ increases,  the PCS is higher but the probability of patients treated over MTD also increases. The median values $\alpha=0.1$     and $\alpha=0.12$   lead to high safety and reasonable reliability when compared to other values. 
In practice, we recommend   starting $\alpha$ at 0.1 or 0.12, and calibrating its value   until desirable simulation results are reached by investigators.

\input{code/table/sen_tol_result}

Next we conduct sensitivity analysis of $K$, the upper limit of the ESS. The results are summarized in   Table \ref{table:sen_K}.   Three   versions of the Hi3+3 design, with   $K=\infty$, $K=12$, and $K=9$ are tested using scenarios 2 and 11 where historical data  has a large sample size at some doses.  In scenario 11, dose 3 has 18 patients in the historical data with the correct MTD, dose 4 receiving no patients. 
Hi3+3 does not assume any information is available on doses 4 and 5, and suffers from the incorrect data (6 DLTs out of 18 patients) on dose 3 from the historical trial.  As a result, Hi3+3 could lead to a low PCS if $K= \infty$; i.e., there is no limitation of the prior ESS. By setting $K$ to 9 or lower, Hi3+3 design will not   borrow too much incorrect information and therefore leads to better performance. Reversely,  Scenario 2 shows that with correct historical data (compared to the truth), the limitation of ESS negatively lead to a slightly lower PCS. Combining observations for both scenarios and our experiences in real-world trials that typically do not enroll a large number of patients at a single dose, we recommend setting $K$ in a range of 9-15 patients.   For example, in Table \ref{table:fixed_scenarios} with $K=9$ Hi3+3 exhibits superior safety in Scenario 6.

\begin{table}[h]
	\centering
	\subtable[Truth and historical data for scenario 2]{
		\centering
		\begin{tabular}{rlllll}
			\hline
			\input{code/table/set2}
		\end{tabular}
	}

	\subtable[Simulation results for scenario 2]{
		\centering
		\begin{tabular}{|p{0.1\textwidth}|p{0.1\textwidth}|p{0.08\textwidth}|p{0.08\textwidth}|p{0.08\textwidth}|p{0.08\textwidth}|p{0.08\textwidth}|p{0.08\textwidth}|p{0.08\textwidth}|p{0.08\textwidth}|}
			\hline
			\input{code/table/sen_K_result1}
		\end{tabular}
	}
	
	\subtable[Truth and historical data for scenario 11]{
		\centering
		\begin{tabular}{rlllll}
			\hline
			\input{code/table/set11}
		\end{tabular}
	}
	
	\subtable[Simulation results for scenario 11]{
		\centering
		\begin{tabular}{|p{0.1\textwidth}|p{0.1\textwidth}|p{0.08\textwidth}|p{0.08\textwidth}|p{0.08\textwidth}|p{0.08\textwidth}|p{0.08\textwidth}|p{0.08\textwidth}|p{0.08\textwidth}|p{0.08\textwidth}|}
			\hline
			\input{code/table/sen_K_result2}
		\end{tabular}
	
	}

\caption{Sensitivity analysis of values of $K$ for scenarios 2 and 11.}
\label{table:sen_K}
	
\end{table}

\section{Efficiency Improvement}
An important advantage of incorporating historical information in phase I clinical trials is that the efficiency of the trial could be improved. This can be quantified by various metrics. For example, using historical data through the Hi3+3 design, the current trial may need a smaller sample size and achieve similar performances when compared to designs not borrowing information. 

We conduct additional simulations to change the sample size of Hi3+3 and evaluate the influence of incorporating historical data on the design with different sample sizes.   Figure \ref{tab:effiency} presents results for three scenarios as examples. Summarizing the results in the three representative scenarios, we find that with a reduction from 30 patients as the sample size for i3+3 to about 24 patients for Hi3+3, the reliability and safety performance of the two designs are comparable. In other words, a reduction of 20\% in sample size is achievable with Hi3+3 when compared to i3+3. Such reduction might be more or less if the historical data is believed to be more or less trustworthy.   For example,   increasing $K$ allows more historical data to be borrowed   and may increase the efficiency of the current trial design if historical data are believed to be trustworthy.    

\newgeometry{top=2cm,bottom=2cm}
\begin{table}[h]
	
	\centering
	\scalebox{0.9}{%
		\subtable[Truth and historical data for scenario 14-1]{
			\centering
			\begin{tabular}{rlllll}
				\hline
				\input{code/table/set_sen_sampsize_1}
			\end{tabular}
		}
	}
	
	\scalebox{0.9}{%
		\subtable[Simulation results for scenario 14-1]{
			\centering
			\begin{tabular}{|p{0.18\textwidth}|p{0.1\textwidth}|p{0.08\textwidth}|p{0.08\textwidth}|p{0.08\textwidth}|p{0.08\textwidth}|p{0.08\textwidth}|p{0.08\textwidth}|p{0.08\textwidth}|p{0.05\textwidth}|}
				\hline
				\input{code/table/sen_sampsize_result1}
			\end{tabular}
			
		}
	}
		
%
%
	\centering
	\scalebox{0.9}{%
		\subtable[Truth and historical data for scenario 14-2]{			
			\centering
			\begin{tabular}{rlllll}
				\hline
				\input{code/table/set_sen_sampsize_2}
			\end{tabular}
		}
	}
	\scalebox{0.9}{%
		\subtable[Simulation results for scenario 14-2]{
			\centering
			\begin{tabular}{|p{0.18\textwidth}|p{0.1\textwidth}|p{0.08\textwidth}|p{0.08\textwidth}|p{0.08\textwidth}|p{0.08\textwidth}|p{0.08\textwidth}|p{0.08\textwidth}|p{0.08\textwidth}|p{0.05\textwidth}|}
				\hline
				\input{code/table/sen_sampsize_result2}
			\end{tabular}
		}
	}
	
	\end{table}
	
	\newgeometry{bottom=2cm}
	\begin{table}[h]
	\centering
	\scalebox{0.9}{%
		\subtable[Truth and historical data for scenario 14-3]{
			\centering
			\begin{tabular}{rlllll}
				\hline
				\input{code/table/set_sen_sampsize_3}
			\end{tabular}
		}
	}
	\scalebox{0.9}{%
		\subtable[Simulation results for scenario 14-3]{
			\centering
			\begin{tabular}{|p{0.18\textwidth}|p{0.1\textwidth}|p{0.08\textwidth}|p{0.08\textwidth}|p{0.08\textwidth}|p{0.08\textwidth}|p{0.08\textwidth}|p{0.08\textwidth}|p{0.08\textwidth}|p{0.05\textwidth}|}
				\hline
				\input{code/table/sen_sampsize_result3}
			\end{tabular}
		}
	}
	
	\caption{Evaluation of efficiency gain for the   Hi3+3 design versus the i3+3 design.  }
	\label{tab:effiency}
\end{table}

\restoregeometry

\section{Discussion}

The Hi3+3 design is a simple design that extends i3+3 by incorporating historical data to improve the efficiency of phase I clinical trials.   The simplicity is highlighted by  the pretabulated decision tables for the candidate doses of the current trial.   
 These tables are easy to understand and guide the decision making for all the enrolled patients throughout the entire trial.   
	
 Hi3+3    ensures safety for patients and trialists via the use of $\alpha$-Tolerability and $K$-Ceiling conditions. These conditions effectively limit the potentially biased and unsafe historical information from being abused for the decision making for current trial patients. This is a critical feature that safeguards the trial design and conduct.   
 Users of Hi3+3 may adjust $\alpha$ and $K$ to calibrate the performance of Hi3+3, including safety.   Once the $\alpha$ value is   restricted   to a relatively low level (e.g. the default value 0.1), the risk of   selecting and allocating doses over MTD   is much reduced.   As shown in the simulation, Hi3+3 remains safe even when the historical data is not compatible with the simulation truth.   
	
	Statistical modeling based on the power prior is extended in Hi3+3 as well. We propose to use dose-dependent ESS (DESS) to enforce better flexibility and safety. 
 A limitation of Hi3+3 is that it could only decide the amount of information borrowed from historical data based on the   toxicity outcomes   of the trial,   while in practice, there may be other baseline   patients characteristics  that could be modeled to better determine the similarities of patient populations in the historical trial and the current trial.   

\bibliographystyle{unsrtnat}
\bibliography{bibfile}

\clearpage \newpage




\begin{table}[h]
	\centering
	\subfigure{
		\begin{minipage}[t]{1\linewidth}
			\centering
			\centerline{\includegraphics[width=1\textwidth]{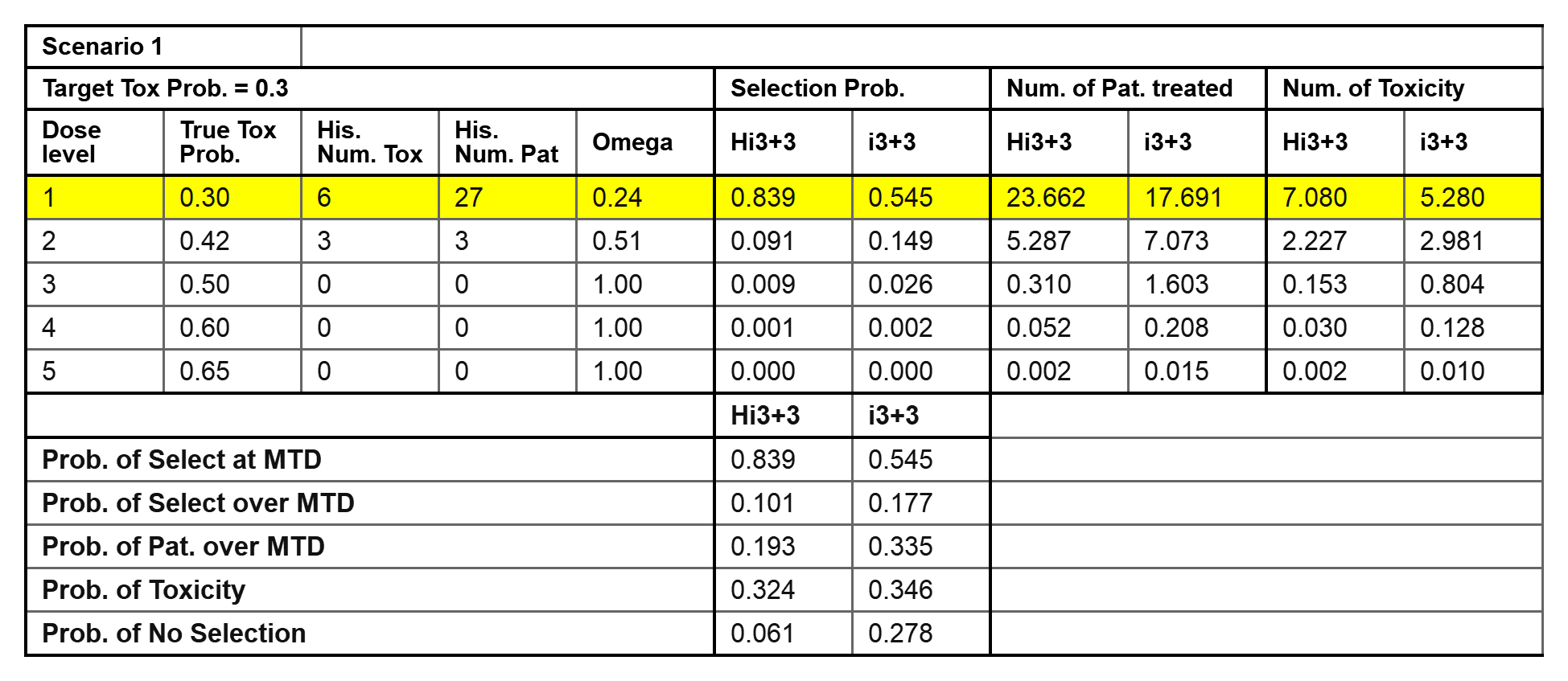}}
		\end{minipage}%
	}%

	\centering
	\subfigure{
		\begin{minipage}[t]{1\linewidth}
			\centering
			\centerline{\includegraphics[width=1\textwidth]{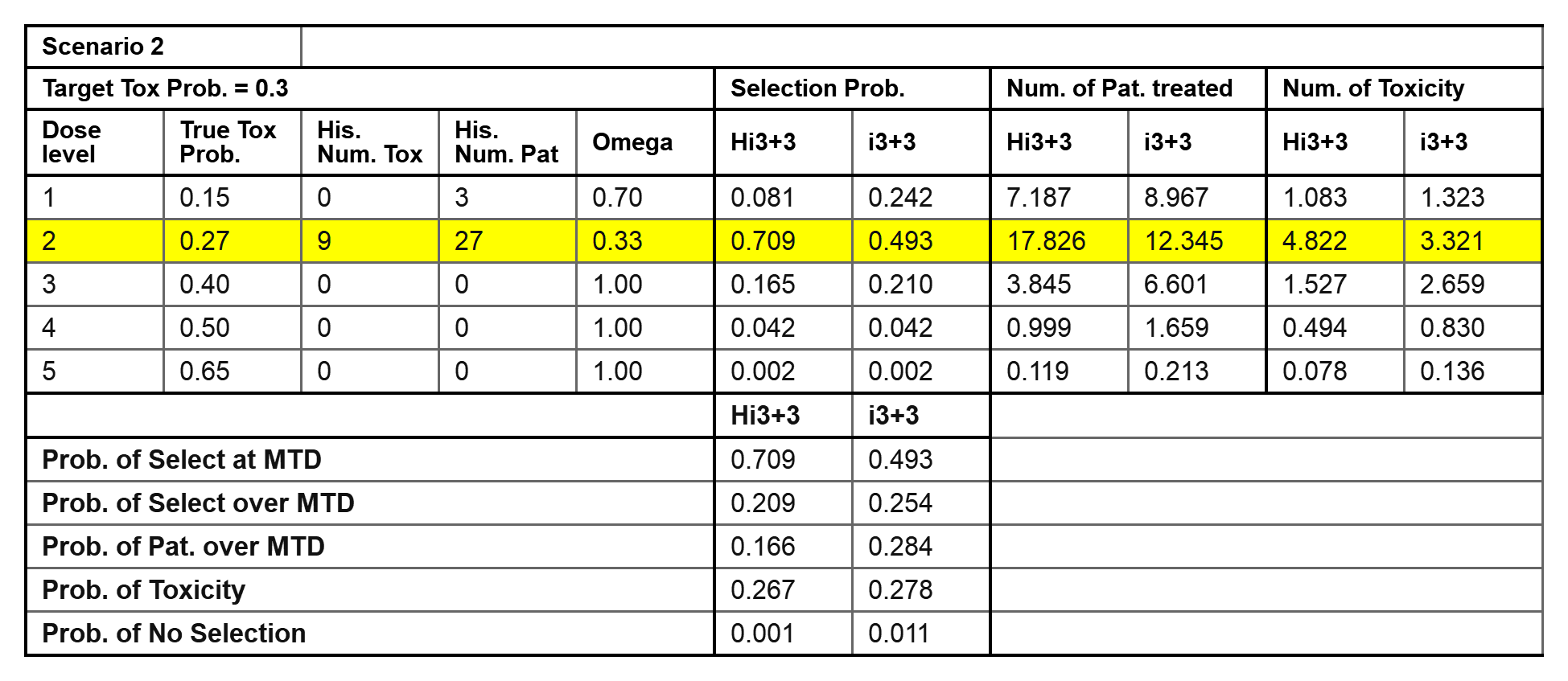}}
		\end{minipage}%
	}%

	\centering	
	\subfigure{
		\begin{minipage}[t]{1\linewidth}
			\centering
			\centerline{\includegraphics[width=1\textwidth]{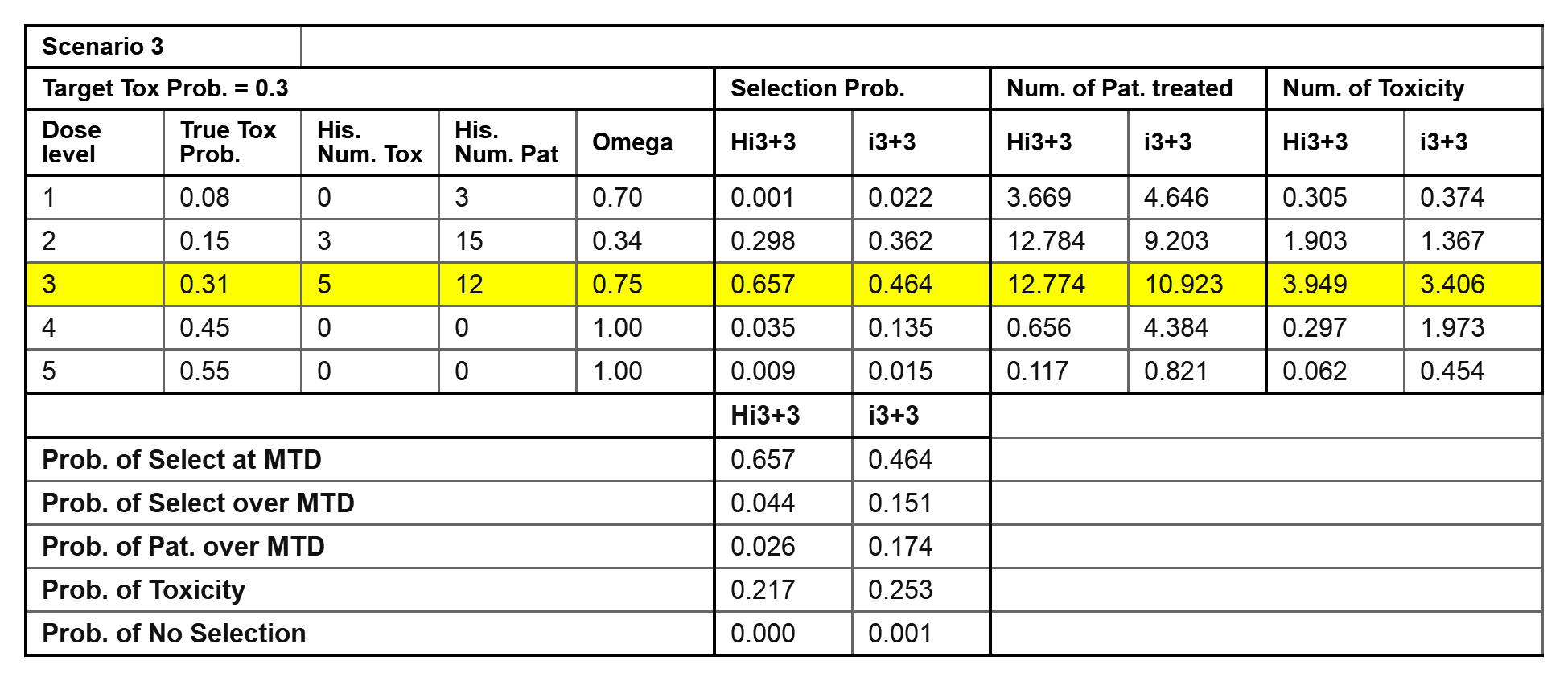}}
		\end{minipage}%
	}%
	
\end{table} 

\addtocounter{table}{-1}       


\begin{table} [h]

\addtocounter{table}{1}

	\centering
	\subfigure{
		\begin{minipage}[t]{1\linewidth}
			\centering
			\centerline{\includegraphics[width=1\textwidth]{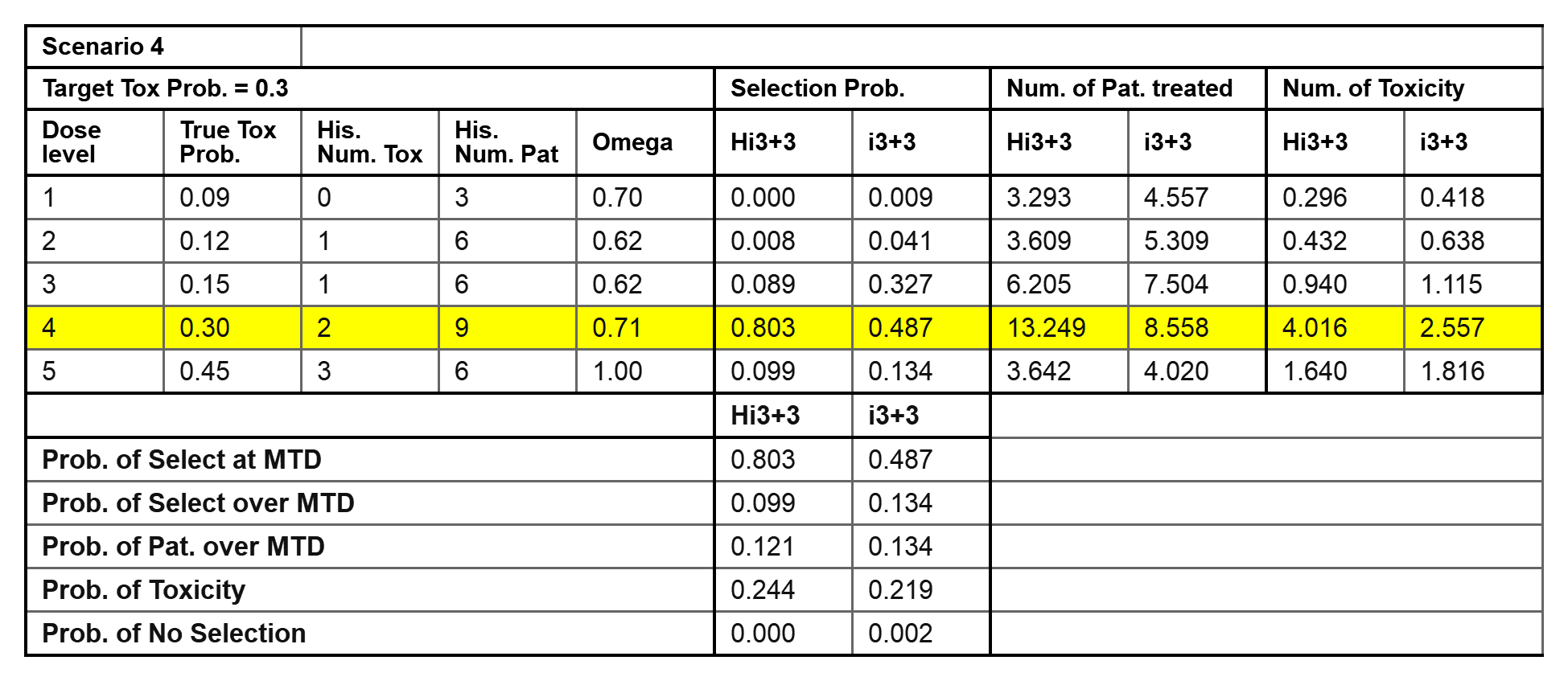}}
		\end{minipage}%
	}%
	
	\centering
	\subfigure{
		\begin{minipage}[t]{1\linewidth}
			\centering
			\centerline{\includegraphics[width=1\textwidth]{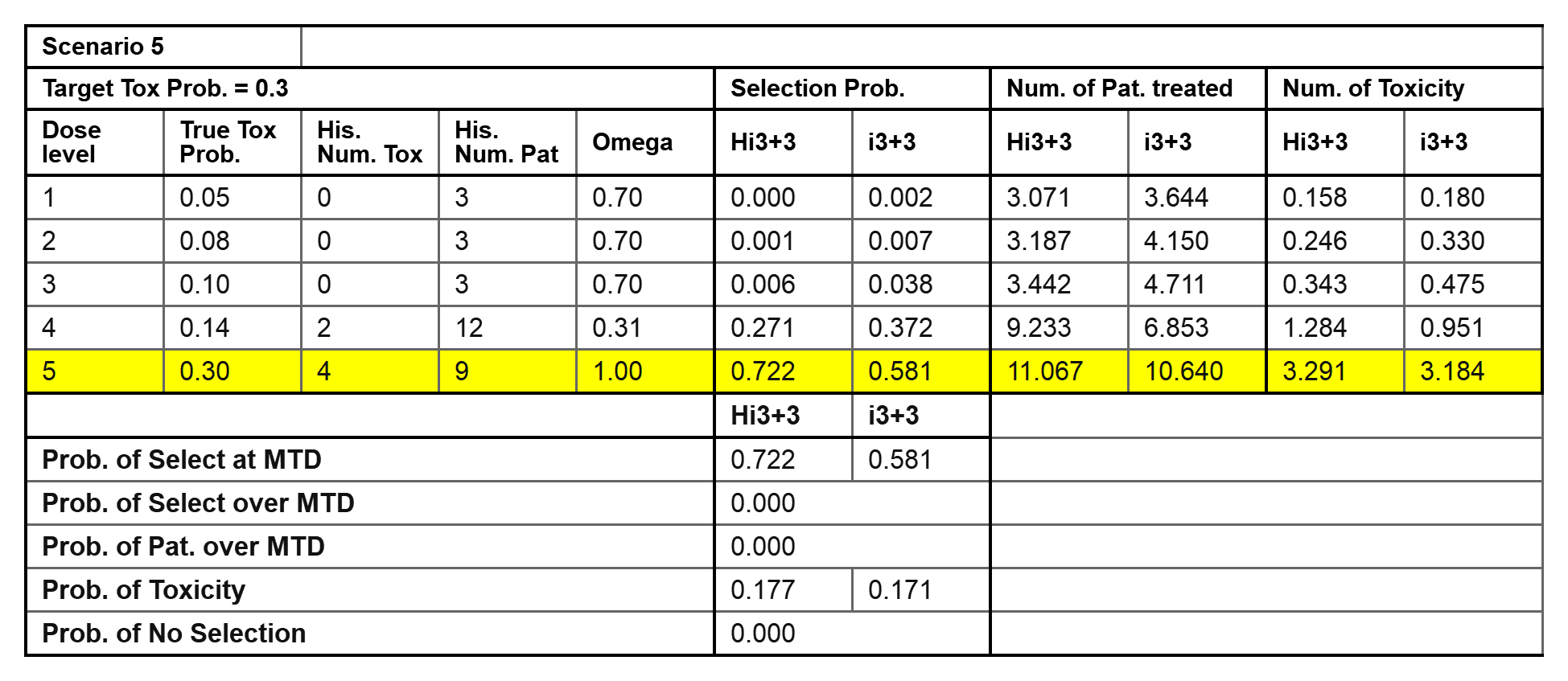}}
		\end{minipage}%
	}%

	\centering
	\subfigure{
		\begin{minipage}[t]{1\linewidth}
			\centering
			\centerline{\includegraphics[width=1\textwidth]{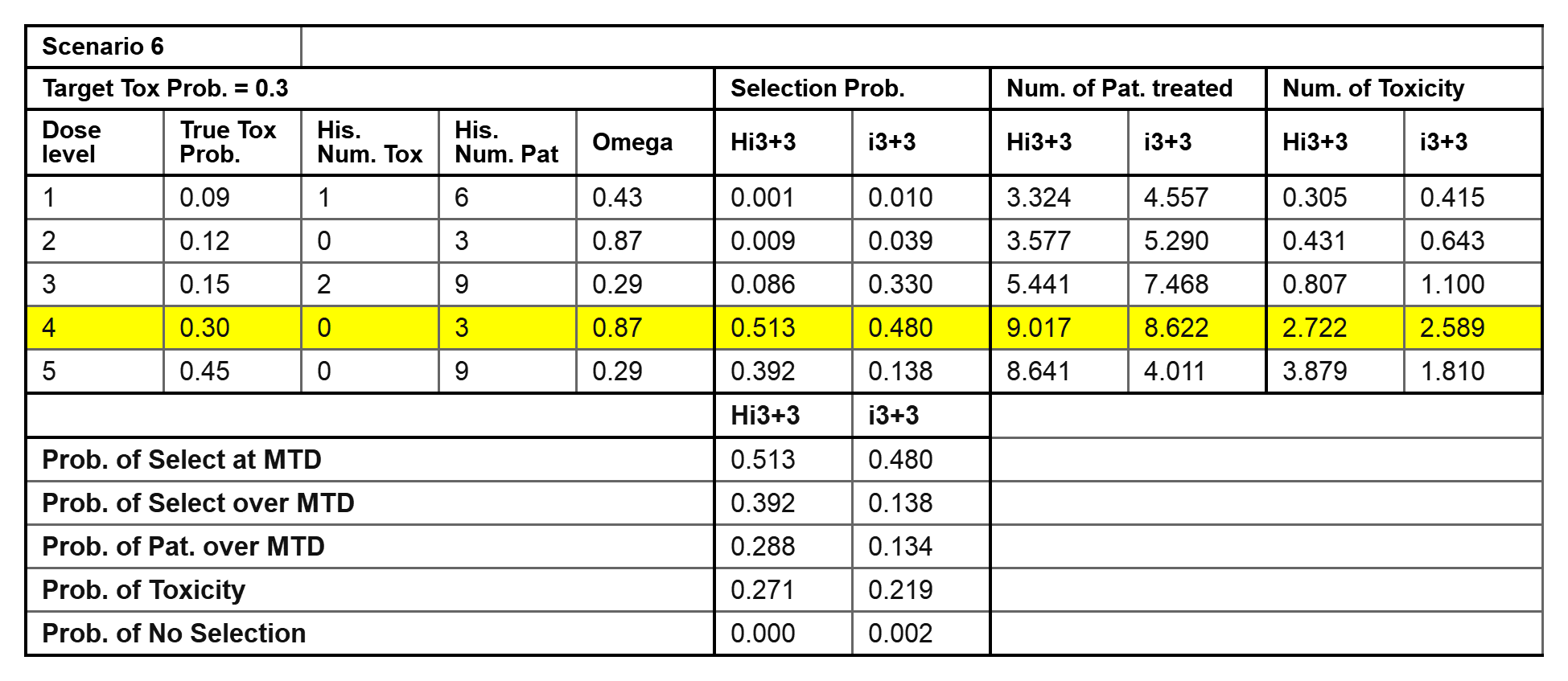}}
		\end{minipage}%
	}%
	
\end{table} 

\addtocounter{table}{-1}       

\begin{table} [h]

\addtocounter{table}{1}

	\centering
	\subfigure{
		\begin{minipage}[t]{1\linewidth}
			\centering
			\centerline{\includegraphics[width=1\textwidth]{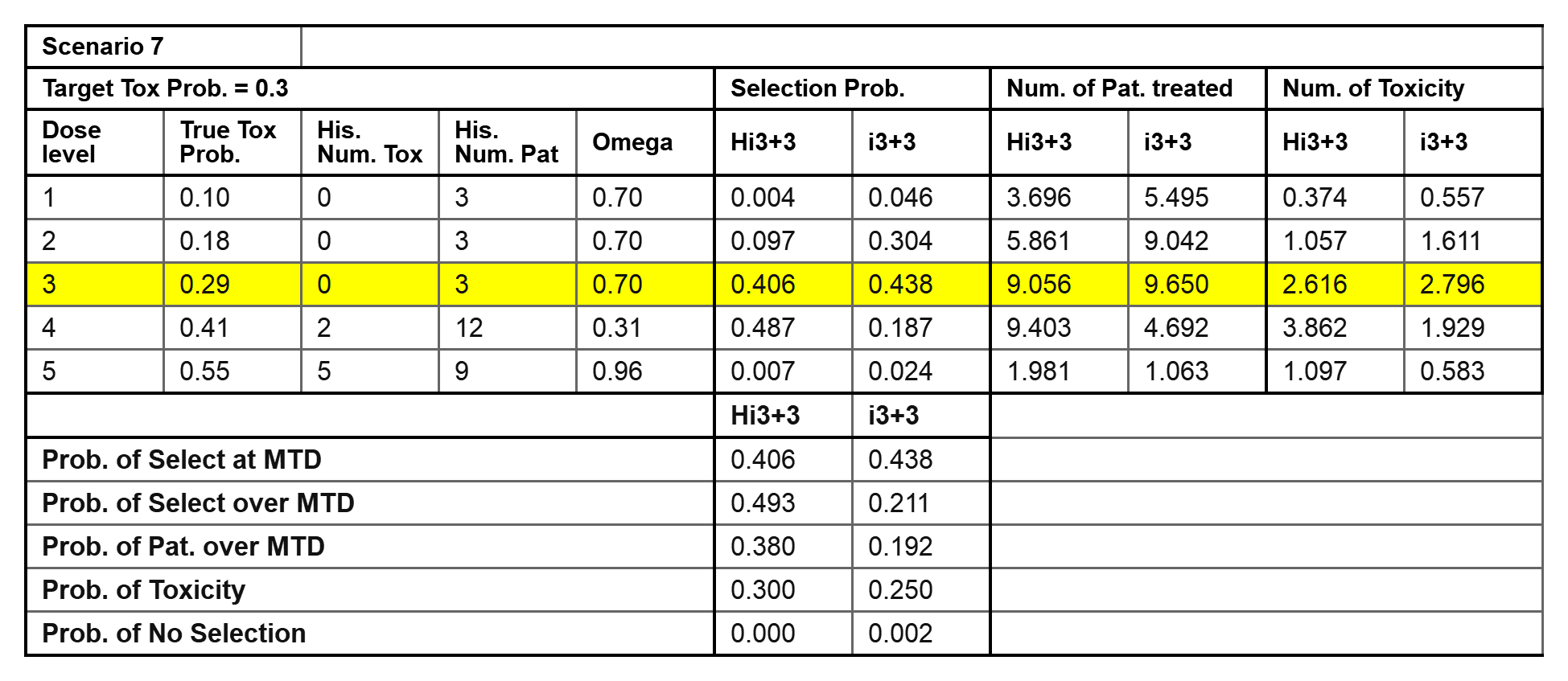}}
		\end{minipage}%
	}%

	\centering
	\subfigure{
		\begin{minipage}[t]{1\linewidth}
			\centering
			\centerline{\includegraphics[width=1\textwidth]{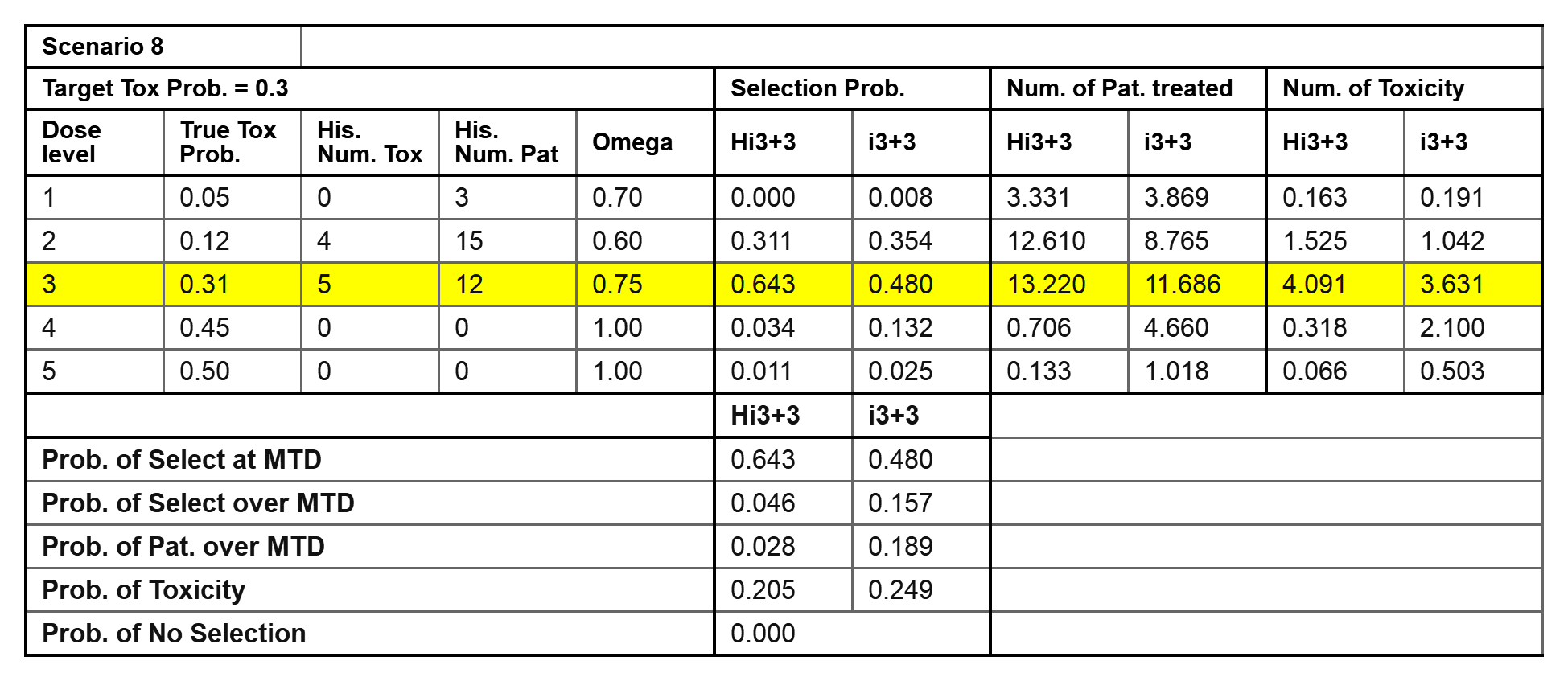}}
		\end{minipage}%
	}%
	
	\centering
	\subfigure{
		\begin{minipage}[t]{1\linewidth}
			\centering
			\centerline{\includegraphics[width=1\textwidth]{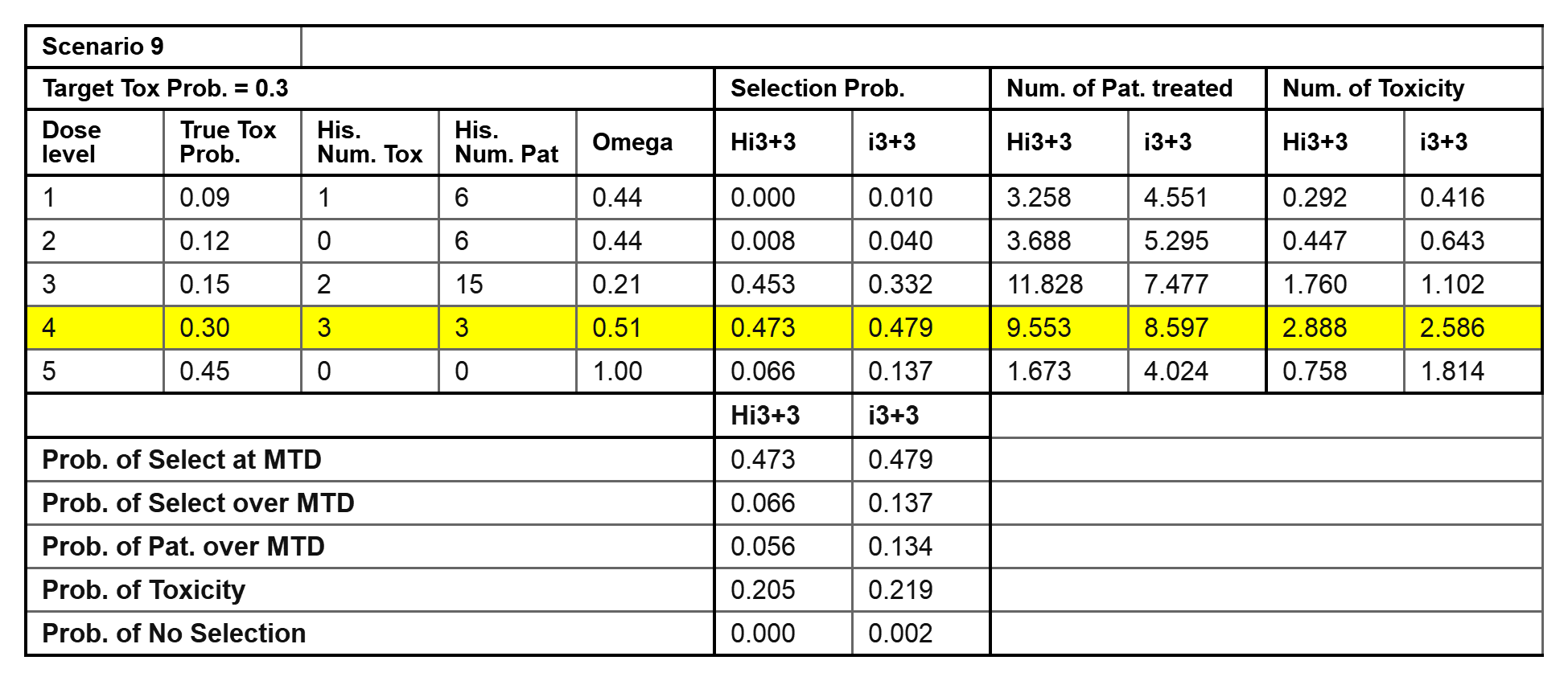}}
		\end{minipage}%
	}%
	
\end{table} 

\addtocounter{table}{-1}       

\begin{table} [h]

\addtocounter{table}{1}
	
	\centering
	\subfigure{
		\begin{minipage}[t]{1\linewidth}
			\centering
			\centerline{\includegraphics[width=1\textwidth]{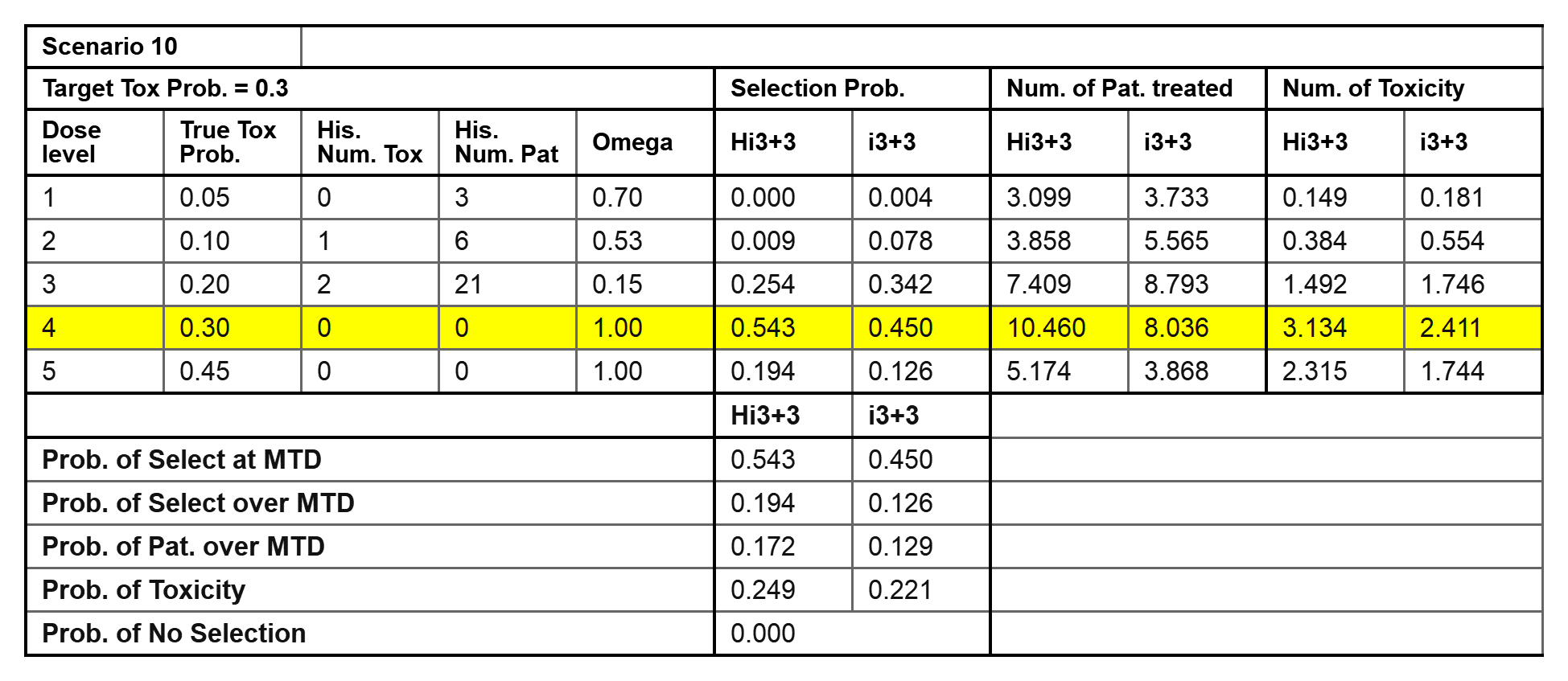}}
		\end{minipage}%
	}%

	\centering
	\subfigure{
		\begin{minipage}[t]{1\linewidth}
			\centering
			\centerline{\includegraphics[width=1\textwidth]{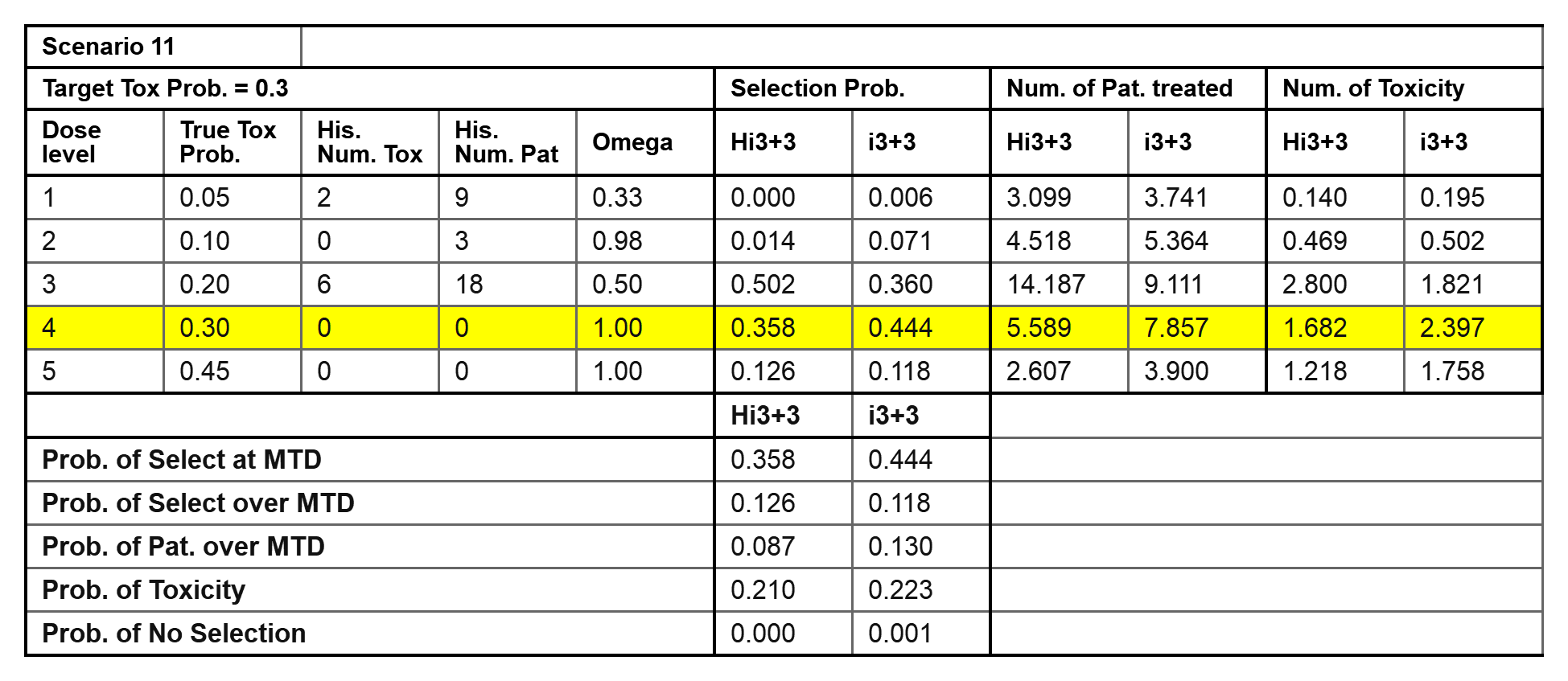}}
		\end{minipage}%
	}%

	\centering
	\subfigure{
		\begin{minipage}[t]{1\linewidth}
			\centering
			\centerline{\includegraphics[width=1\textwidth]{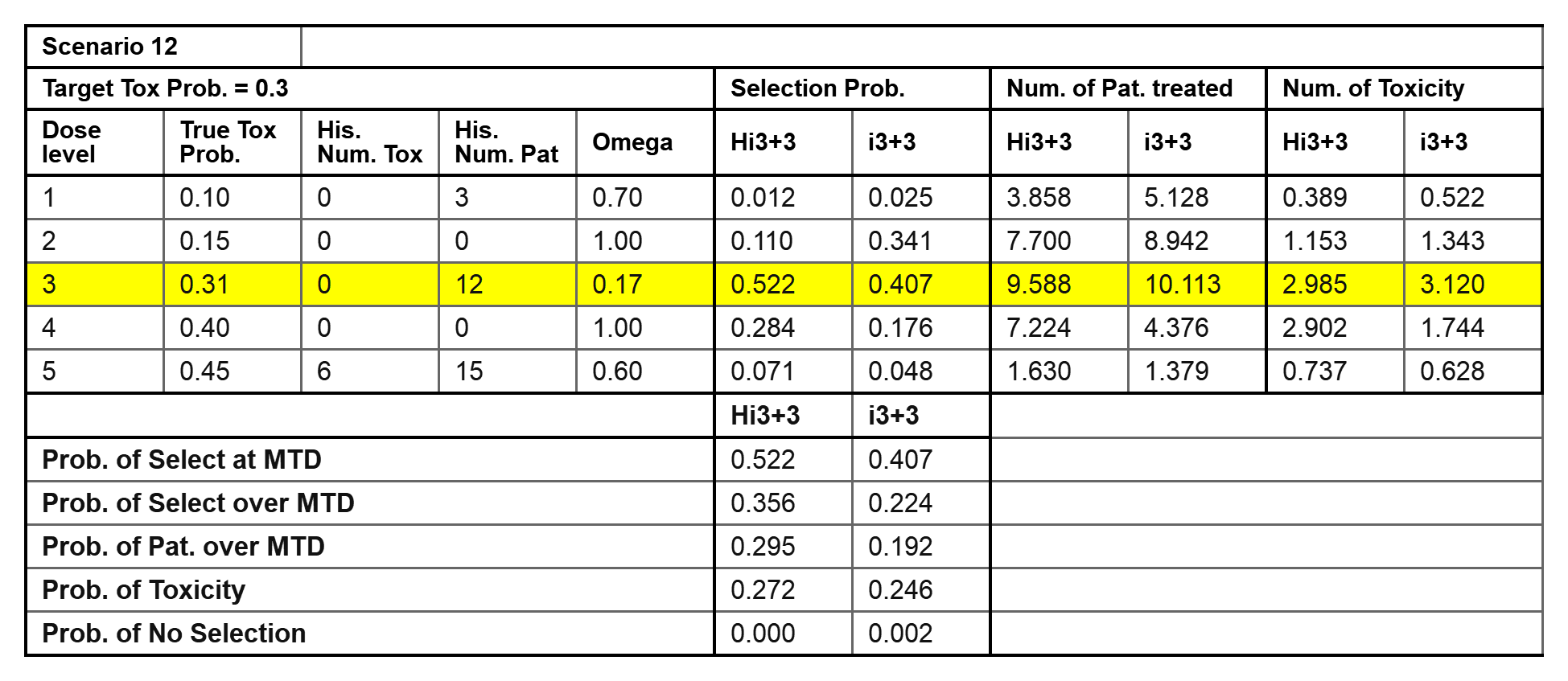}}
		\end{minipage}%
	}%
	
\end{table} 

\addtocounter{table}{-1}

\begin{table}

\addtocounter{table}{1}
	
	\centering
	\subfigure{
		\begin{minipage}[t]{1\linewidth}
			\centering
			\centerline{\includegraphics[width=1\textwidth]{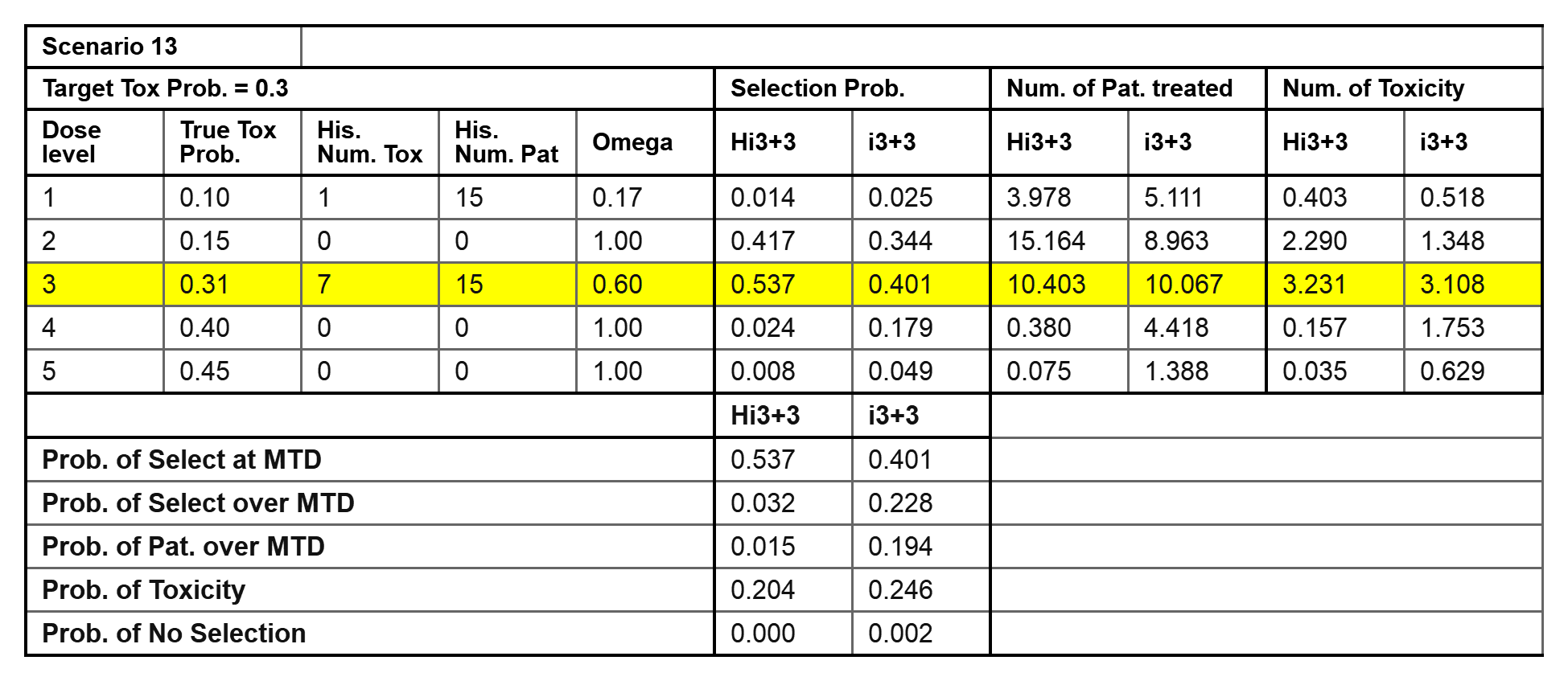}}
		\end{minipage}%
	}%

	\centering
	\caption{  Simulation results of the 13 fixed scenarios comparing Hi3+3 and i3+3.  }
	\label{table:fixed_scenarios}
\end{table}

\clearpage
\newpage
\restoregeometry
\section*{Appendix A}

In the {\bf Safety rules}, the two posterior probabilities are calculated under the posterior distribution $f'(p_d|M_0,M,W) = beta(x_d + a_d^\star + 1 - a_0, n_d - x_d + m_d - a_d^\star + 1 - b_0)$ using the isotonic-transformed prior \eqref{eq:power2} as the prior for $p_d$ and the binomial likelihood for the current data $(x_d, n_d)$.   Note that  $f'(p_d|M_0,M,W) \propto beta(p_d|a_d^\star +1- a_0, m_d-a_d^\star +1- b_0) \cdot bin(x_d|n_d,p_d)$, where $beta(p_d|a_d^\star +1- a_0, m_d-a_d^\star +1- b_0)$ is the isotonic transformed power prior based on the initial $beta(1,1)$ prior instead of the $beta(a_0, b_0)$ prior in \eqref{eq:power1}.   Therefore, we have the term $(1-a_0)$ and $(1-b_0)$ in the expression. The use of $beta(1,1)$ as a prior leads to more conservative inference since $beta(1,1)$ assumes {\it a priori} the prior mean for each dose is $0.5$ (usually the MTD target $p_T$ is less than 0.5) with an effective sample size of 2.   Therefore, $beta(1,1)$ tends to skew the posterior estimates of dose toxicity probability towards 0.5 and encourages stopping or dose exclusion in both rules. 

\section*{Appendix B}

\renewcommand{\thetable}{B\arabic{table}}

\setcounter{table}{0}

  Full results comparing Hi3+3, i3+3, iBOIN, and iBOIN$_R$ are summarized in the two tables below, for the fixed and random scenarios, respectively. In general, the three designs borrowing historical data exhibit superior performance when the historical data are compatible with the truth. This is expected of course. Across the three designs Hi3+3, iBOIN, and iBOIN$_R$, their performance is largely similar. Depending on the assumptions of the designs, their performance varies by scenarios. However, Hi3+3 shows the promised ``robustness'' in that when the historical data are not compatible with the scenario truth, it still can limit the patients on doses that are overly toxic.  



\begin{table}[h]
	\centering
	\subfigure{
		\begin{minipage}[t]{1\linewidth}
			\centering
			\centerline{\includegraphics[width=1\textwidth]{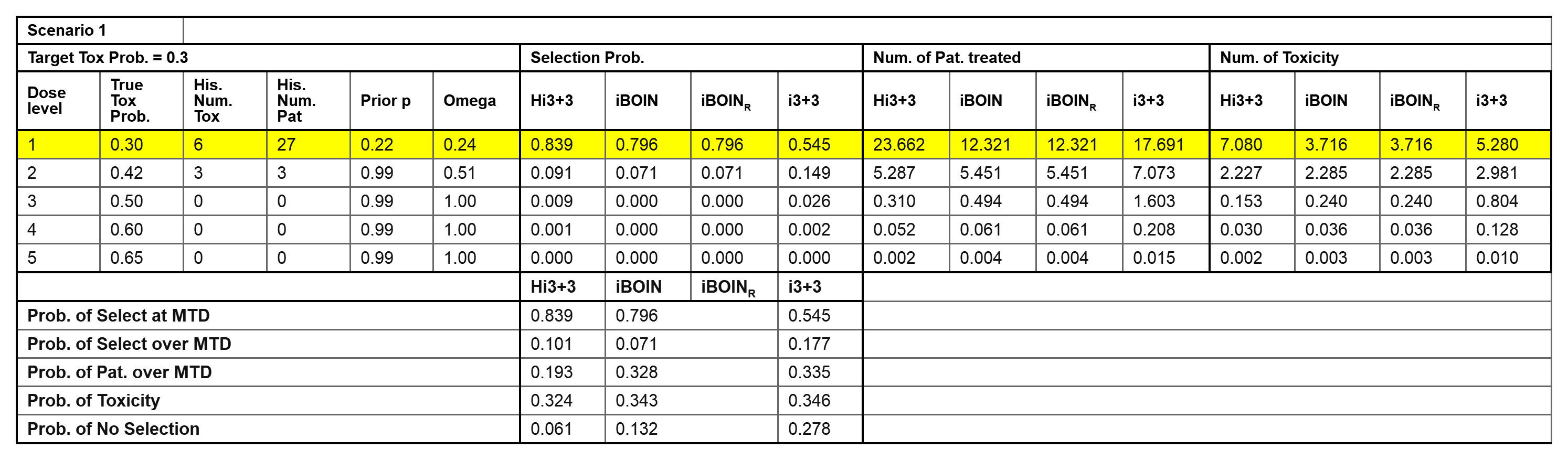}}
		\end{minipage}%
	}%

	\centering
	\subfigure{
		\begin{minipage}[t]{1\linewidth}
			\centering
			\centerline{\includegraphics[width=1\textwidth]{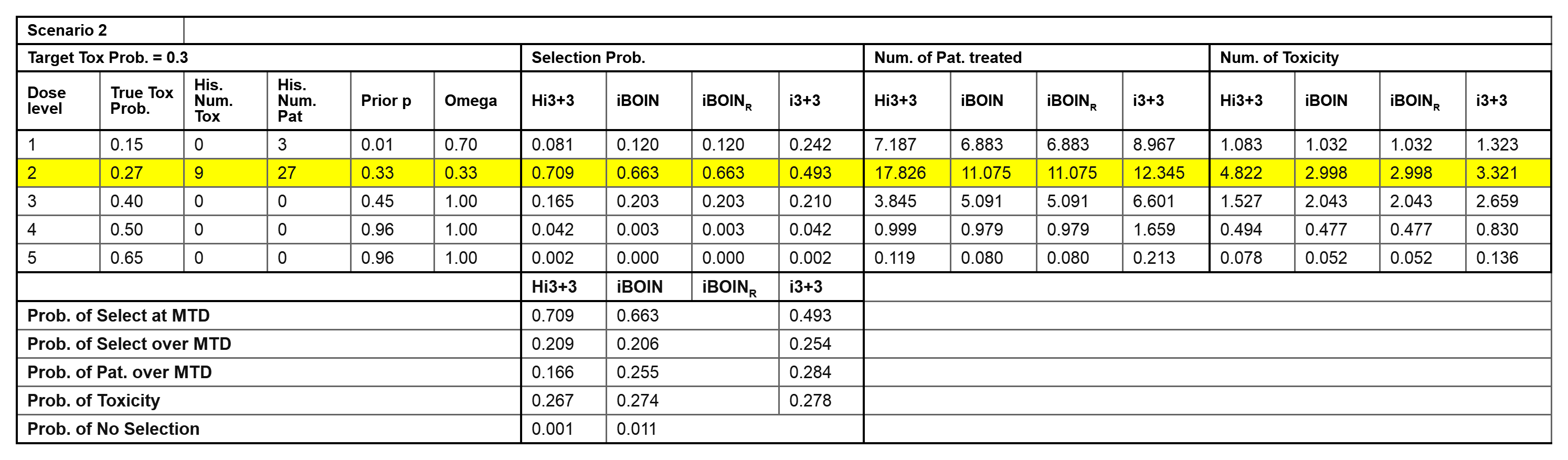}}
		\end{minipage}%
	}%

	\centering	
	\subfigure{
		\begin{minipage}[t]{1\linewidth}
			\centering
			\centerline{\includegraphics[width=1\textwidth]{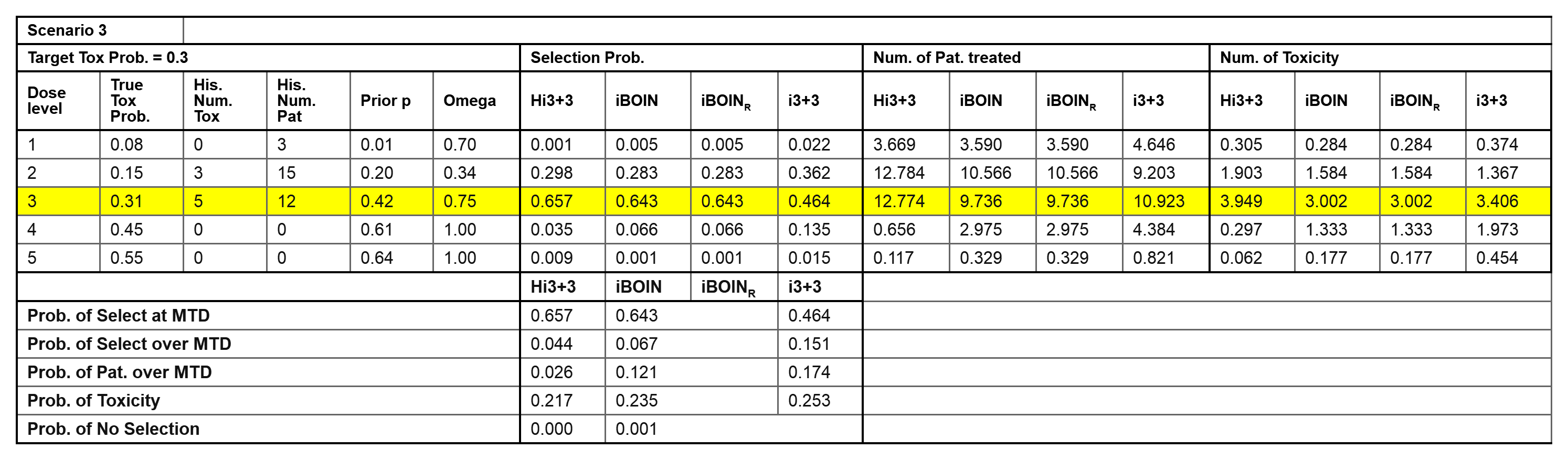}}
		\end{minipage}%
	}%
	
\end{table} 

\addtocounter{table}{-1}       


\begin{figure} [h]
	
	\addtocounter{table}{1}

	\centering
	\subfigure{
		\begin{minipage}[t]{1\linewidth}
			\centering
			\centerline{\includegraphics[width=1\textwidth]{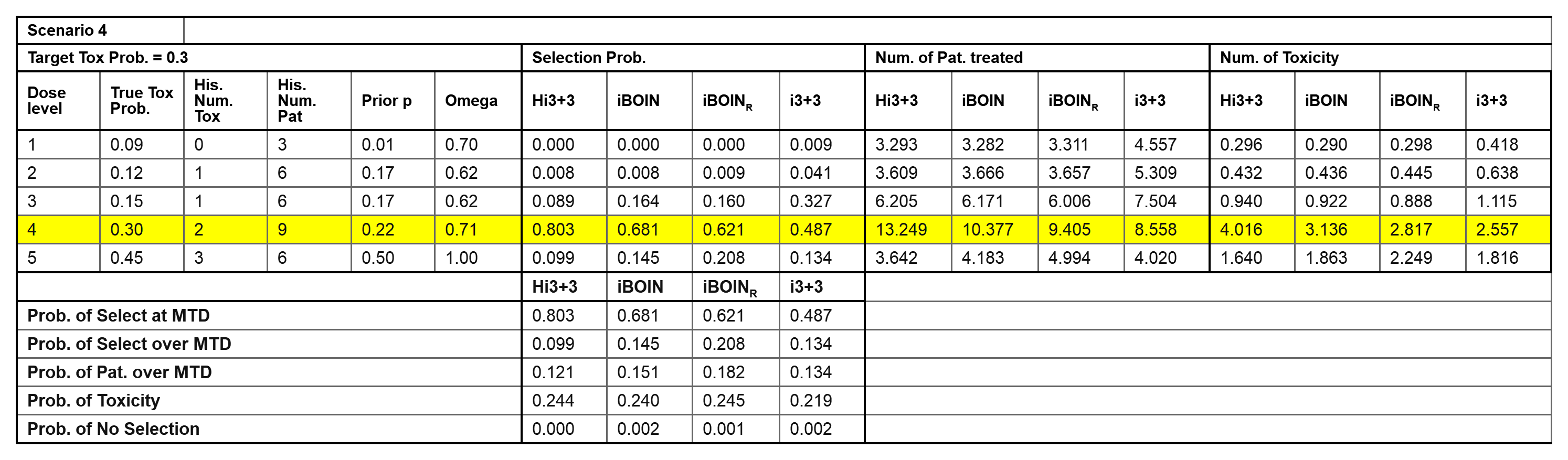}}
		\end{minipage}%
	}%
	
	\centering
	\subfigure{
		\begin{minipage}[t]{1\linewidth}
			\centering
			\centerline{\includegraphics[width=1\textwidth]{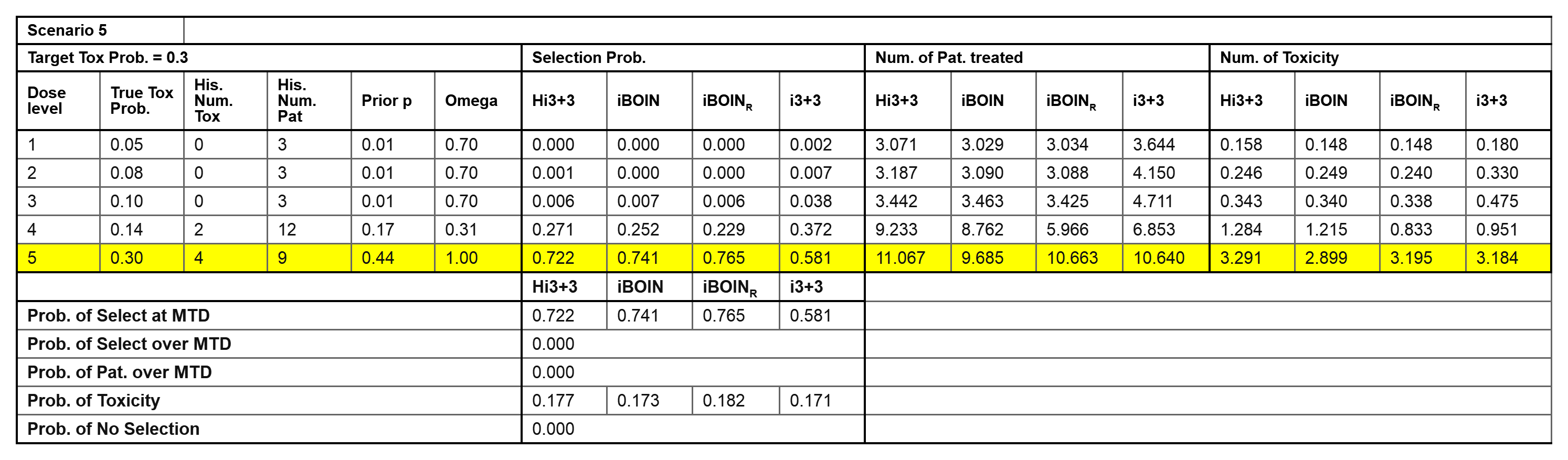}}
		\end{minipage}%
	}%

	\centering
	\subfigure{
		\begin{minipage}[t]{1\linewidth}
			\centering
			\centerline{\includegraphics[width=1\textwidth]{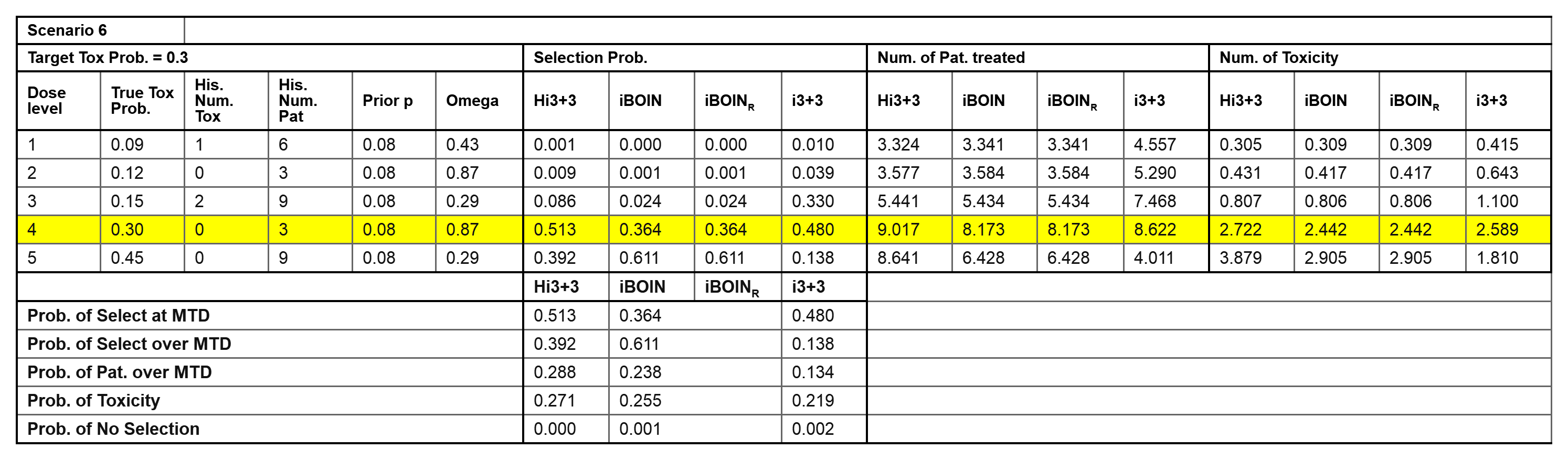}}
		\end{minipage}%
	}%

	\centering
	\subfigure{
		\begin{minipage}[t]{1\linewidth}
			\centering
			\centerline{\includegraphics[width=1\textwidth]{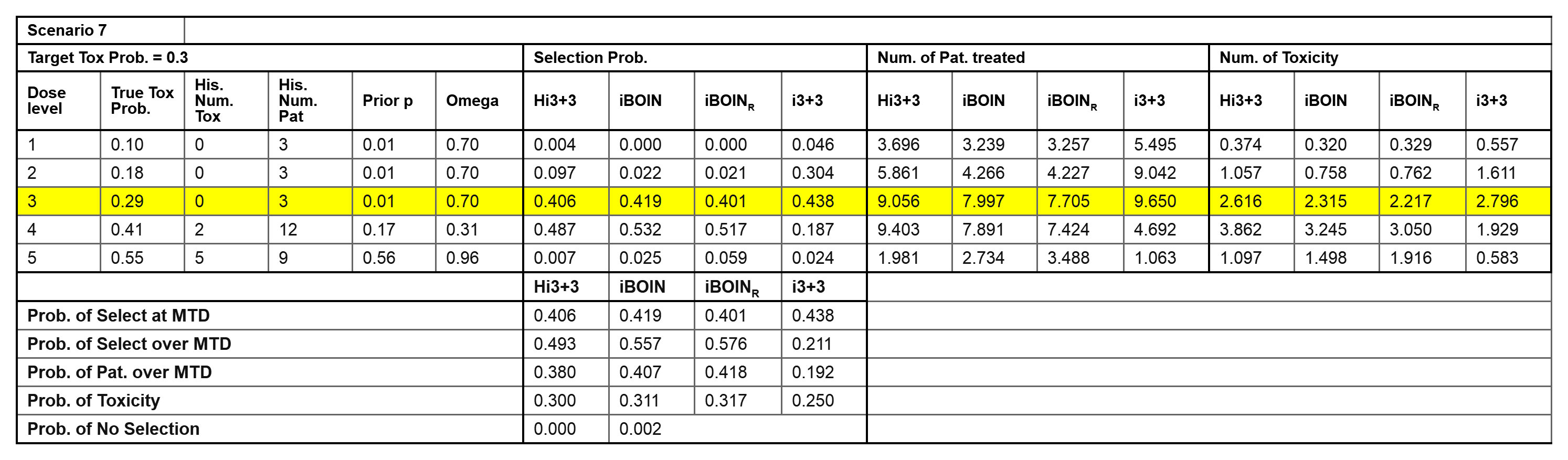}}
		\end{minipage}%
	}%

\end{figure} 

\addtocounter{table}{-1}       

\begin{table} [h]
	
	\addtocounter{table}{1}

	\centering
	\subfigure{
		\begin{minipage}[t]{1\linewidth}
			\centering
			\centerline{\includegraphics[width=1\textwidth]{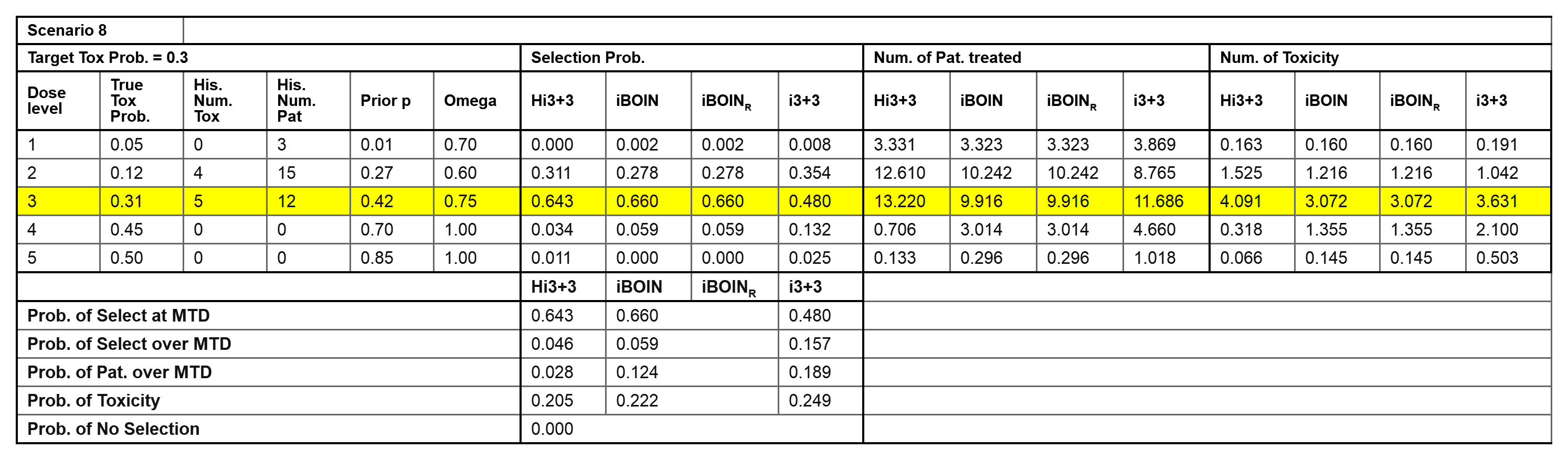}}
		\end{minipage}%
	}%
	
	\centering
	\subfigure{
		\begin{minipage}[t]{1\linewidth}
			\centering
			\centerline{\includegraphics[width=1\textwidth]{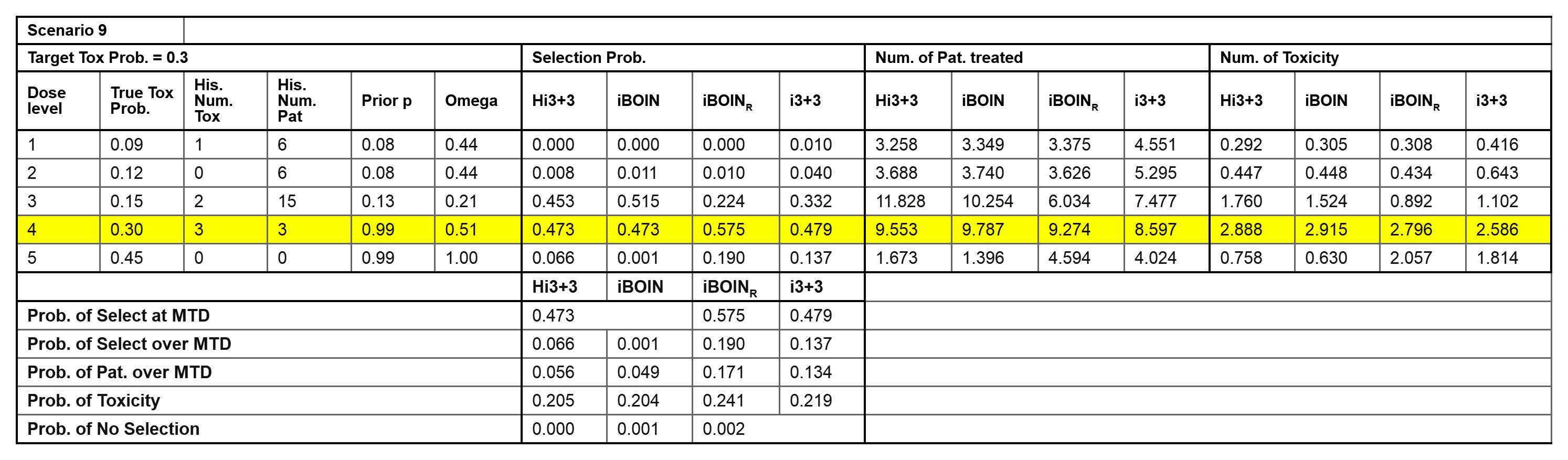}}
		\end{minipage}%
	}%

	\centering
	\subfigure{
		\begin{minipage}[t]{1\linewidth}
			\centering
			\centerline{\includegraphics[width=1\textwidth]{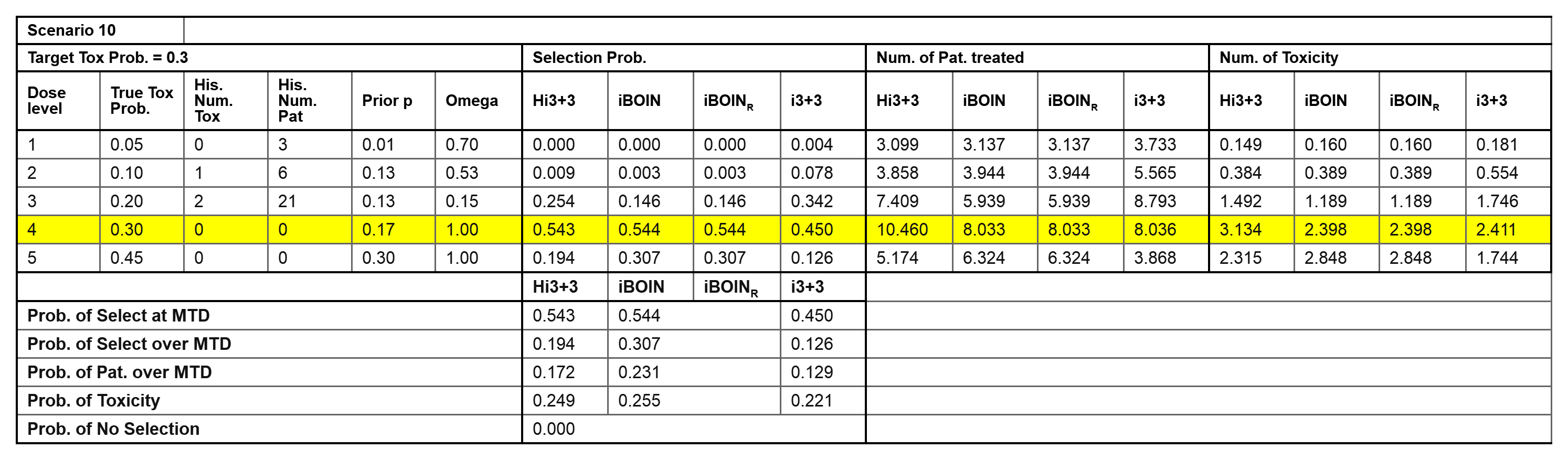}}
		\end{minipage}%
	}%
	
	\centering
	\subfigure{
		\begin{minipage}[t]{1\linewidth}
			\centering
			\centerline{\includegraphics[width=1\textwidth]{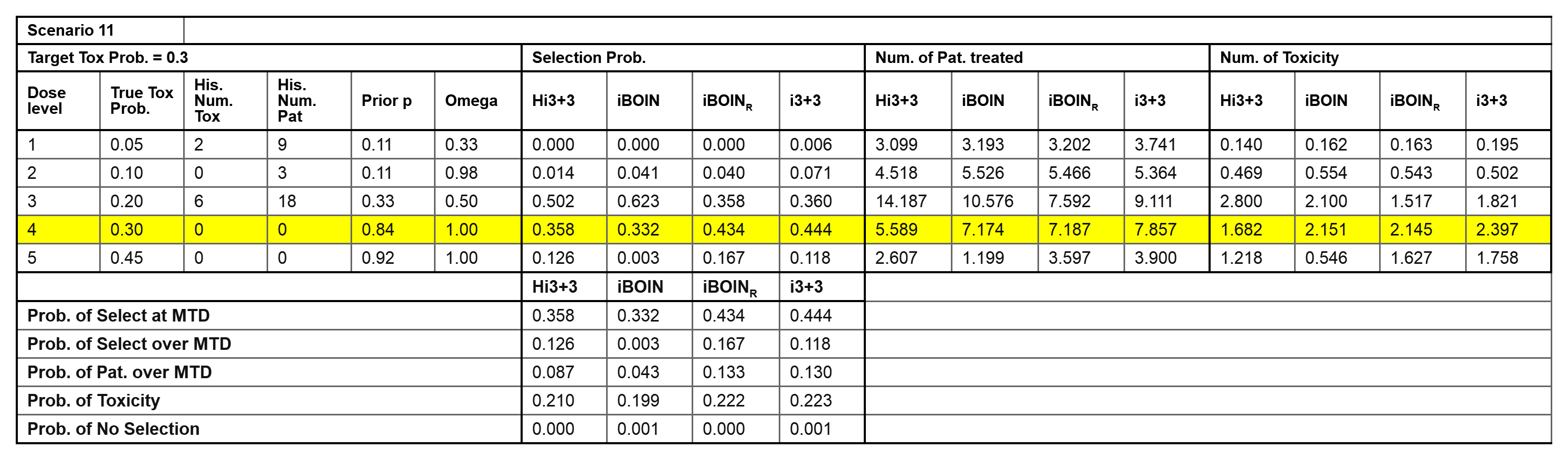}}
		\end{minipage}%
	}%

\end{table} 

\addtocounter{table}{-1}

\begin{table}
	
	\addtocounter{table}{1}
	
	\centering
	\subfigure{
		\begin{minipage}[t]{1\linewidth}
			\centering
			\centerline{\includegraphics[width=1\textwidth]{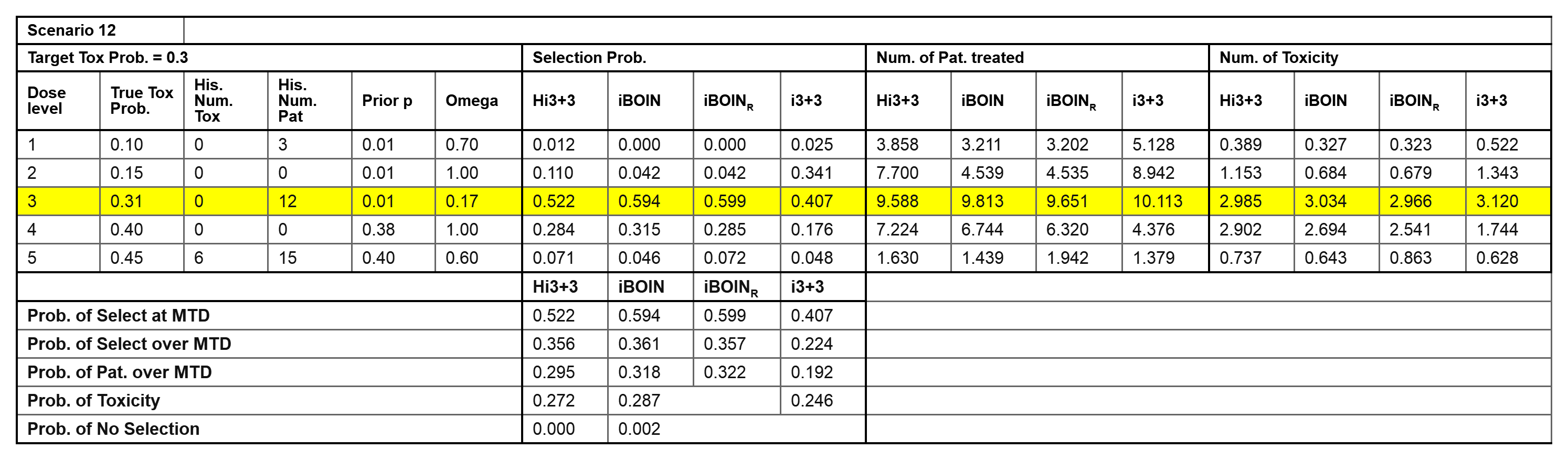}}
		\end{minipage}%
	}%
	
	\centering
	\subfigure{
		\begin{minipage}[t]{1\linewidth}
			\centering
			\centerline{\includegraphics[width=1\textwidth]{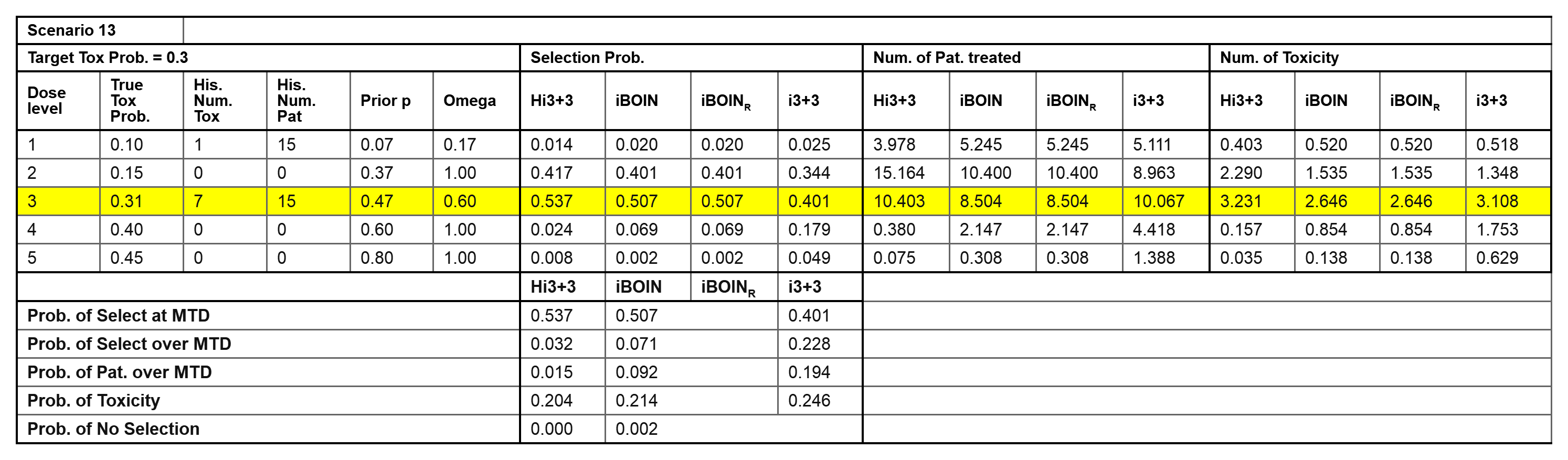}}
		\end{minipage}%
	}%

	\centering
	\caption{  Simulation results of the 13 fixed scenarios comparing Hi3+3, iBOIN, iBOIN$_R$, and i3+3.  }
	\label{table:fixed_scenarios_appendix}
\end{table}

\clearpage
\newpage

\input{code/table/result_random_table}

\restoregeometry

\end{document}

%% file: YJ.tex
\usepackage{graphicx}

\usepackage[dvipsnames,usenames]{color}

\usepackage{amsmath}                        

\definecolor{DarkGreen}{rgb}{0.5,0.8,0.6}   
\definecolor{RGBblack}{rgb}{0.0,0.0,0.0}    

\newcommand{\hei}[1]{\color{black}{#1} \color{black} }
\definecolor{grau}{rgb}{0.8,0.8,0.8}

\newcommand{\chen}[1]{\color{orange}}

\def\hei{\color{black}}

%% file: code/table/result_random_table2.tex
\begin{table}[ht]
\centering
\begin{tabular}{rll}
  \hline
 & Hi3+3 & i3+3 \\ 
  \hline
\#Correct Sel. of MTD & $0.693_{(0.213)}$ & $0.561_{(0.134)}$ \\ 
  \#Sel. over MTD & $0.076_{(0.119)}$ & $0.089_{(0.092)}$ \\ 
  \#Pat over MTD & $0.086_{(0.108)}$ & $0.128_{(0.111)}$ \\ 
  \# Tox & $0.261_{(0.050)}$ & $0.265_{(0.053)}$ \\ 
  \#None Sel. & $0.045_{(0.113)}$ & $0.078_{(0.124)}$ \\ 
   \hline
\end{tabular}
\caption{Results of 1000 random scenarios} 
\label{table:random_result}
\end{table}

%% file: code/table/sen_tol_result.tex
\begin{table}[ht]
\centering
\begin{tabular}{|p{0.1\textwidth}|p{0.1\textwidth}|p{0.1\textwidth}|p{0.1\textwidth}|p{0.08\textwidth}|p{0.08\textwidth}|p{0.08\textwidth}|p{0.08\textwidth}|p{0.08\textwidth}|}
  \hline
 Hi3+3 & \#Correct Sel. of MTD & \#Sel. over MTD & \#Sel. under MTD & \#None Sel. & \#Pat. at MTD & \#Pat. over MTD & \#Pat. under MTD & \#Tox \\ 
   \hline
$\alpha$=0.05 & 0.680 & 0.108 & 0.211 & 0.001 & 0.339 & 0.092 & 0.568 & 0.218 \\ 
  $\alpha$=0.08 & 0.754 & 0.095 & 0.151 & 0.000 & 0.358 & 0.100 & 0.542 & 0.225 \\ 
  $\alpha$=0.10 & 0.816 & 0.080 & 0.104 & 0.000 & 0.445 & 0.116 & 0.439 & 0.248 \\ 
  $\alpha$=0.12 & 0.815 & 0.100 & 0.085 & 0.000 & 0.470 & 0.123 & 0.406 & 0.251 \\ 
  $\alpha$=0.15 & 0.838 & 0.092 & 0.070 & 0.000 & 0.496 & 0.123 & 0.381 & 0.254 \\ 
  $\alpha$=0.20 & 0.841 & 0.101 & 0.058 & 0.000 & 0.495 & 0.130 & 0.375 & 0.254 \\ 
   \hline
i3+3 & 0.488 & 0.130 & 0.380 & 0.002 & 0.277 & 0.137 & 0.586 & 0.217 \\ 
   \hline
\end{tabular}
\caption{Sensitivity analysis of Hi3+3 with different $\alpha$ values in the $\alpha$-Tolerability condition.} 
\label{table:sen_tol}
\end{table}

%% file: code/table/set2.tex
 & 1 & 2 & 3 & 4 & 5 \\ 
  \hline
True Prob & 0.15 & 0.27 & 0.40 & 0.50 & 0.65 \\ 
  history x &  0 &  9 &  0 &  0 &  0 \\ 
  history n &  3 & 27 &  0 &  0 &  0 \\ 
   \hline

%% file: code/table/sen_K_result1.tex
 Hi3+3 & \#Correct Sel. of MTD & \#Sel. over MTD & \#Sel. under MTD & \#None Sel. & \#Pat. at MTD & \#Pat. over MTD & \#Pat. under MTD & \#Tox \\ 
   \hline
$K=\infty$ & 0.898 & 0.056 & 0.046 & 0.000 & 0.723 & 0.026 & 0.251 & 0.248 \\ 
  $K=12$ & 0.760 & 0.148 & 0.092 & 0.000 & 0.615 & 0.130 & 0.254 & 0.263 \\ 
  $K=9$ & 0.699 & 0.198 & 0.103 & 0.000 & 0.591 & 0.154 & 0.254 & 0.266 \\ 
  $K=6$ & 0.642 & 0.240 & 0.118 & 0.000 & 0.517 & 0.226 & 0.256 & 0.277 \\ 
   \hline
i3+3 & 0.493 & 0.250 & 0.247 & 0.010 & 0.416 & 0.271 & 0.313 & 0.273 \\ 
   \hline

%% file: code/table/set11.tex
 & 1 & 2 & 3 & 4 & 5 \\ 
  \hline
True Prob & 0.05 & 0.10 & 0.20 & 0.30 & 0.45 \\ 
  history x &  2 &  0 &  6 &  0 &  0 \\ 
  history n &  9 &  3 & 18 &  0 &  0 \\ 
   \hline

%% file: code/table/sen_K_result2.tex
 Hi3+3 & \#Correct Sel. of MTD & \#Sel. over MTD & \#Sel. under MTD & \#None Sel. & \#Pat. at MTD & \#Pat. over MTD & \#Pat. under MTD & \#Tox \\ 
   \hline
$K=\infty$ & 0.212 & 0.106 & 0.682 & 0.000 & 0.106 & 0.043 & 0.851 & 0.191 \\ 
  $K=12$ & 0.340 & 0.112 & 0.548 & 0.000 & 0.170 & 0.078 & 0.751 & 0.207 \\ 
  $K=9$ & 0.358 & 0.126 & 0.516 & 0.000 & 0.186 & 0.087 & 0.727 & 0.210 \\ 
  $K=6$ & 0.392 & 0.148 & 0.460 & 0.000 & 0.246 & 0.116 & 0.638 & 0.224 \\ 
   \hline
i3+3 & 0.444 & 0.118 & 0.437 & 0.001 & 0.262 & 0.130 & 0.608 & 0.223 \\ 
   \hline

%% file: code/table/set_sen_sampsize_1.tex
 & 1 & 2 & 3 & 4 & 5 \\ 
  \hline
True Prob & 0.08 & 0.15 & 0.31 & 0.48 & 0.55 \\ 
  history x & 0.00 & 3.00 & 5.00 & 0.00 & 0.00 \\ 
  history n & 3.00 & 15.00 & 12.00 & 0.00 & 0.00 \\ 
   \hline

%% file: code/table/sen_sampsize_result1.tex
 Hi3+3 & \#Correct Sel. of MTD & \#Sel. over MTD & \#Sel. under MTD & \#None Sel. & \#Pat. at MTD & \#Pat. over MTD & \#Pat. under MTD & \#Tox \\ 
   \hline
samplesize=15 & 0.621 & 0.000 & 0.379 & 0.000 & 0.296 & 0.000 & 0.704 & 0.179 \\ 
  samplesize=18 & 0.697 & 0.013 & 0.290 & 0.001 & 0.357 & 0.004 & 0.639 & 0.193 \\ 
  samplesize=21 & 0.671 & 0.023 & 0.306 & 0.001 & 0.388 & 0.009 & 0.603 & 0.203 \\ 
  samplesize=24 & 0.642 & 0.032 & 0.325 & 0.000 & 0.401 & 0.015 & 0.584 & 0.209 \\ 
  samplesize=27 & 0.657 & 0.035 & 0.308 & 0.000 & 0.418 & 0.019 & 0.562 & 0.214 \\ 
  samplesize=30 & 0.660 & 0.037 & 0.303 & 0.000 & 0.424 & 0.025 & 0.551 & 0.219 \\ 
  samplesize=33 & 0.673 & 0.035 & 0.292 & 0.000 & 0.468 & 0.031 & 0.502 & 0.228 \\ 
  samplesize=36 & 0.683 & 0.039 & 0.278 & 0.000 & 0.471 & 0.035 & 0.494 & 0.230 \\ 
  samplesize=39 & 0.680 & 0.037 & 0.283 & 0.000 & 0.477 & 0.038 & 0.485 & 0.234 \\ 
  samplesize=42 & 0.679 & 0.039 & 0.281 & 0.000 & 0.481 & 0.039 & 0.480 & 0.234 \\ 
  samplesize=45 & 0.683 & 0.038 & 0.279 & 0.000 & 0.487 & 0.040 & 0.472 & 0.237 \\ 
   \hline
i3+3 &  &  &  &  &  &  &  &  \\ 
   \hline
samplesize=30 & 0.493 & 0.124 & 0.382 & 0.001 & 0.373 & 0.165 & 0.461 & 0.254 \\ 
   \hline

%% file: code/table/set_sen_sampsize_2.tex
 & 1 & 2 & 3 & 4 & 5 \\ 
  \hline
True Prob & 0.08 & 0.15 & 0.31 & 0.48 & 0.55 \\ 
  history x & 0.00 & 0.00 & 1.00 & 3.00 & 3.00 \\ 
  history n & 3.00 & 3.00 & 9.00 & 9.00 & 6.00 \\ 
   \hline

%% file: code/table/sen_sampsize_result2.tex
 Hi3+3 & \#Correct Sel. of MTD & \#Sel. over MTD & \#Sel. under MTD & \#None Sel. & \#Pat. at MTD & \#Pat. over MTD & \#Pat. under MTD & \#Tox \\ 
   \hline
samplesize=15 & 0.479 & 0.278 & 0.242 & 0.001 & 0.284 & 0.078 & 0.638 & 0.201 \\ 
  samplesize=18 & 0.488 & 0.326 & 0.185 & 0.000 & 0.376 & 0.157 & 0.467 & 0.248 \\ 
  samplesize=21 & 0.551 & 0.326 & 0.123 & 0.000 & 0.345 & 0.265 & 0.390 & 0.282 \\ 
  samplesize=24 & 0.566 & 0.319 & 0.114 & 0.000 & 0.361 & 0.280 & 0.359 & 0.291 \\ 
  samplesize=27 & 0.631 & 0.262 & 0.107 & 0.000 & 0.379 & 0.279 & 0.341 & 0.295 \\ 
  samplesize=30 & 0.655 & 0.241 & 0.104 & 0.000 & 0.392 & 0.278 & 0.330 & 0.298 \\ 
  samplesize=33 & 0.683 & 0.229 & 0.088 & 0.000 & 0.415 & 0.278 & 0.307 & 0.302 \\ 
  samplesize=36 & 0.718 & 0.208 & 0.074 & 0.000 & 0.457 & 0.274 & 0.269 & 0.306 \\ 
  samplesize=39 & 0.741 & 0.179 & 0.081 & 0.000 & 0.483 & 0.268 & 0.248 & 0.310 \\ 
  samplesize=42 & 0.743 & 0.174 & 0.083 & 0.000 & 0.497 & 0.267 & 0.236 & 0.312 \\ 
  samplesize=45 & 0.771 & 0.146 & 0.084 & 0.000 & 0.501 & 0.281 & 0.218 & 0.319 \\ 
   \hline
i3+3 &  &  &  &  &  &  &  &  \\ 
   \hline
samplesize=30 & 0.493 & 0.124 & 0.382 & 0.001 & 0.373 & 0.165 & 0.461 & 0.254 \\ 
   \hline

%% file: code/table/set_sen_sampsize_3.tex
 & 1 & 2 & 3 & 4 & 5 \\ 
  \hline
True Prob & 0.08 & 0.15 & 0.31 & 0.48 & 0.55 \\ 
  history x & 2.00 & 1.00 & 6.00 & 0.00 & 0.00 \\ 
  history n & 9.00 & 9.00 & 12.00 & 0.00 & 0.00 \\ 
   \hline

%% file: code/table/sen_sampsize_result3.tex
 Hi3+3 & \#Correct Sel. of MTD & \#Sel. over MTD & \#Sel. under MTD & \#None Sel. & \#Pat. at MTD & \#Pat. over MTD & \#Pat. under MTD & \#Tox \\ 
   \hline
samplesize=15 & 0.496 & 0.156 & 0.348 & 0.000 & 0.284 & 0.078 & 0.639 & 0.201 \\ 
  samplesize=18 & 0.495 & 0.193 & 0.312 & 0.000 & 0.312 & 0.114 & 0.574 & 0.221 \\ 
  samplesize=21 & 0.526 & 0.192 & 0.281 & 0.000 & 0.331 & 0.138 & 0.531 & 0.235 \\ 
  samplesize=24 & 0.525 & 0.202 & 0.273 & 0.000 & 0.343 & 0.156 & 0.500 & 0.246 \\ 
  samplesize=27 & 0.571 & 0.184 & 0.245 & 0.000 & 0.408 & 0.194 & 0.398 & 0.272 \\ 
  samplesize=30 & 0.584 & 0.170 & 0.246 & 0.000 & 0.420 & 0.195 & 0.385 & 0.276 \\ 
  samplesize=33 & 0.597 & 0.161 & 0.242 & 0.000 & 0.430 & 0.200 & 0.369 & 0.280 \\ 
  samplesize=36 & 0.619 & 0.145 & 0.236 & 0.000 & 0.438 & 0.198 & 0.364 & 0.281 \\ 
  samplesize=39 & 0.627 & 0.139 & 0.235 & 0.000 & 0.448 & 0.196 & 0.357 & 0.281 \\ 
  samplesize=42 & 0.643 & 0.111 & 0.246 & 0.000 & 0.458 & 0.187 & 0.356 & 0.282 \\ 
  samplesize=45 & 0.661 & 0.103 & 0.235 & 0.000 & 0.470 & 0.182 & 0.348 & 0.282 \\ 
   \hline
i3+3 &  &  &  &  &  &  &  &  \\ 
   \hline
samplesize=30 & 0.499 & 0.119 & 0.381 & 0.001 & 0.376 & 0.162 & 0.462 & 0.254 \\ 
   \hline

%% file: code/table/result_random_table.tex
\begin{table}[ht]
\centering
\begin{tabular}{rllll}
  \hline
 & Hi3+3 & iBOIN & iBOIN$_R$ & i3+3 \\ 
  \hline
\#Correct Sel. of MTD & $0.693_{(0.213)}$ & $0.700_{(0.210)}$ & $0.698_{(0.197)}$ & $0.561_{(0.134)}$ \\ 
  \#Sel. over MTD & $0.076_{(0.119)}$ & $0.078_{(0.132)}$ & $0.091_{(0.137)}$ & $0.089_{(0.092)}$ \\ 
  \#Pat over MTD & $0.086_{(0.108)}$ & $0.127_{(0.126)}$ & $0.134_{(0.127)}$ & $0.128_{(0.111)}$ \\ 
  \# Tox & $0.261_{(0.050)}$ & $0.269_{(0.054)}$ & $0.271_{(0.053)}$ & $0.265_{(0.053)}$ \\ 
  \#None Sel. & $0.045_{(0.113)}$ & $0.040_{(0.064)}$ & $0.040_{(0.064)}$ & $0.078_{(0.124)}$ \\ 
   \hline
\end{tabular}
\caption{Results of 1000 random scenarios, Hi3+3, iBOIN, iBOIN$_R$, i3+3} 
\label{table:random_result_appendix}
\end{table}

%% file: version6.bbl
\begin{thebibliography}{20}
\providecommand{\natexlab}[1]{#1}
\providecommand{\url}[1]{\texttt{#1}}
\expandafter\ifx\csname urlstyle\endcsname\relax
  \providecommand{\doi}[1]{doi: #1}\else
  \providecommand{\doi}{doi: \begingroup \urlstyle{rm}\Url}\fi

\bibitem[Liu et~al.(2020)Liu, Wang, and Ji]{liu2020i3+}
Meizi Liu, Sue-Jane Wang, and Yuan Ji.
\newblock {The i3+ 3 design for phase I clinical trials}.
\newblock \emph{Journal of Biopharmaceutical Statistics}, 30\penalty0
  (2):\penalty0 294--304, 2020.

\bibitem[Storer(1989)]{storer1989design}
Barry~E Storer.
\newblock {Design and analysis of phase I clinical trials}.
\newblock \emph{Biometrics}, pages 925--937, 1989.

\bibitem[O'Quigley et~al.(1990)O'Quigley, Pepe, and Fisher]{o1990continual}
John O'Quigley, Margaret Pepe, and Lloyd Fisher.
\newblock {Continual reassessment method: a practical design for phase 1
  clinical trials in cancer}.
\newblock \emph{Biometrics}, pages 33--48, 1990.

\bibitem[Neuenschwander et~al.(2008)Neuenschwander, Branson, and
  Gsponer]{neuenschwander2008critical}
Beat Neuenschwander, Michael Branson, and Thomas Gsponer.
\newblock {Critical aspects of the Bayesian approach to phase I cancer trials}.
\newblock \emph{Statistics in medicine}, 27\penalty0 (13):\penalty0 2420--2439,
  2008.

\bibitem[Ji et~al.(2010)Ji, Liu, Li, and Nebiyou~Bekele]{ji2010modified}
Yuan Ji, Ping Liu, Yisheng Li, and B~Nebiyou~Bekele.
\newblock {A modified toxicity probability interval method for dose-finding
  trials}.
\newblock \emph{Clinical trials}, 7\penalty0 (6):\penalty0 653--663, 2010.

\bibitem[{Beat Neuenschwander and Satrajit Roychoudhury and Heinz
  Schmidli}(2016)]{Neuen2016codata}
{Beat Neuenschwander and Satrajit Roychoudhury and Heinz Schmidli}.
\newblock On the use of co-data in clinical trials.
\newblock \emph{Statistics in Biopharmaceutical Research}, 8\penalty0
  (3):\penalty0 345--354, 2016.
\newblock \doi{10.1080/19466315.2016.1174149}.
\newblock URL \url{https://doi.org/10.1080/19466315.2016.1174149}.

\bibitem[Schmidli et~al.(2014)Schmidli, Gsteiger, Roychoudhury, O'Hagan,
  Spiegelhalter, and Neuenschwander]{schmidli2014robust}
Heinz Schmidli, Sandro Gsteiger, Satrajit Roychoudhury, Anthony O'Hagan, David
  Spiegelhalter, and Beat Neuenschwander.
\newblock {Robust meta-analytic-predictive priors in clinical trials with
  historical control information}.
\newblock \emph{Biometrics}, 70\penalty0 (4):\penalty0 1023--1032, 2014.

\bibitem[Zhou et~al.(2020)Zhou, Lee, Wang, Bailey, and
  Yuan]{zhou2020incorporating}
Yanhong Zhou, J~Jack Lee, Shunguang Wang, Stuart Bailey, and Ying Yuan.
\newblock {Incorporating historical information to improve phase I clinical
  trial designs}.
\newblock \emph{arXiv preprint arXiv:2004.12972}, 2020.

\bibitem[Liu et~al.(2015)Liu, Pan, Xia, Huang, and Yuan]{liu2015bridging}
Suyu Liu, Haitao Pan, Jielai Xia, Qin Huang, and Ying Yuan.
\newblock {Bridging continual reassessment method for phase I clinical trials
  in different ethnic populations}.
\newblock \emph{Statistics in medicine}, 34\penalty0 (10):\penalty0 1681--1694,
  2015.

\bibitem[Li and Yuan(2020)]{li2020pa}
Yimei Li and Ying Yuan.
\newblock {PA-CRM: A continuous reassessment method for pediatric phase I
  oncology trials with concurrent adult trials}.
\newblock \emph{Biometrics}, 2020.

\bibitem[Guo et~al.(2017)Guo, Wang, Yang, Lynn, and Ji]{guo2017bayesian}
Wentian Guo, Sue-Jane Wang, Shengjie Yang, Henry Lynn, and Yuan Ji.
\newblock {A Bayesian interval dose-finding design addressingOckham's razor:
  mTPI-2}.
\newblock \emph{Contemporary clinical trials}, 58:\penalty0 23--33, 2017.

\bibitem[Yan et~al.(2017)Yan, Mandrekar, and Yuan]{yan2017keyboard}
Fangrong Yan, Sumithra~J Mandrekar, and Ying Yuan.
\newblock {Keyboard: a novel Bayesian toxicity probability interval design for
  phase I clinical trials}.
\newblock \emph{Clinical Cancer Research}, 23\penalty0 (15):\penalty0
  3994--4003, 2017.

\bibitem[Ivanova et~al.(2007)Ivanova, Flournoy, and
  Chung]{ivanova2007cumulative}
Anastasia Ivanova, Nancy Flournoy, and Yeonseung Chung.
\newblock {Cumulative cohort design for dose-finding}.
\newblock \emph{Journal of Statistical Planning and Inference}, 137\penalty0
  (7):\penalty0 2316--2327, 2007.

\bibitem[liu(2015)]{liu2015bayesian}
{Bayesian optimal interval designs for phase I clinical trials}, author={Liu,
  Suyu and Yuan, Ying}.
\newblock \emph{Journal of the Royal Statistical Society: Series C: Applied
  Statistics}, pages 507--523, 2015.

\bibitem[Morita et~al.(2008)Morita, Thall, and
  M{\"u}ller]{morita2008determining}
Satoshi Morita, Peter~F Thall, and Peter M{\"u}ller.
\newblock {Determining the effective sample size of a parametric prior}.
\newblock \emph{Biometrics}, 64\penalty0 (2):\penalty0 595--602, 2008.

\bibitem[Ibrahim et~al.(2015)Ibrahim, Chen, Gwon, and Chen]{ibrahim2015power}
Joseph~G Ibrahim, Ming-Hui Chen, Yeongjin Gwon, and Fang Chen.
\newblock {The power prior: theory and applications}.
\newblock \emph{Statistics in medicine}, 34\penalty0 (28):\penalty0 3724--3749,
  2015.

\bibitem[Duan et~al.(2006)Duan, Ye, and Smith]{duan2006evaluating}
Yuyan Duan, Keying Ye, and Eric~P Smith.
\newblock Evaluating water quality using power priors to incorporate historical
  information.
\newblock \emph{Environmetrics: The Official Journal of the International
  Environmetrics Society}, 17\penalty0 (1):\penalty0 95--106, 2006.

\bibitem[Robertson et~al.(1988)Robertson, Wright, and
  Dykstra]{robertson1988order}
T.~Robertson, F.T. Wright, and R.~Dykstra.
\newblock \emph{Order Restricted Statistical Inference}.
\newblock Probability and Statistics Series. Wiley, 1988.
\newblock ISBN 9780471917878.
\newblock URL \url{https://books.google.com/books?id=sqZfQgAACAAJ}.

\bibitem[Bacchetti(1989)]{bacchetti1989additive}
Peter Bacchetti.
\newblock Additive isotonic models.
\newblock \emph{Journal of the American Statistical Association}, 84\penalty0
  (405):\penalty0 289--294, 1989.

\bibitem[Clertant et~al.(2017)Clertant, O’Quigley,
  et~al.]{clertant2017semiparametric}
M~Clertant, J~O’Quigley, et~al.
\newblock {Semiparametric dose finding methods}.
\newblock \emph{Journal of the Royal Statistical Society Series B}, 79\penalty0
  (5):\penalty0 1487--1508, 2017.

\end{thebibliography}
